\documentclass[12pt, draftclsnofoot, onecolumn]{IEEEtran}
\usepackage{amssymb}
\usepackage{amsmath}
\usepackage{cite}
\usepackage{url}
\usepackage{xcolor}
\usepackage{cite,graphicx,amsmath,amssymb}
\usepackage{subfigure}
\usepackage{citesort}
\usepackage{fancyhdr}
\usepackage{mdwmath}
\usepackage{mdwtab}
\usepackage{caption}
\usepackage{amsthm}
\usepackage{setspace}
\usepackage{amsmath}
\usepackage{algorithm}
\usepackage{algorithmic}
\newtheorem{remark}{Remark}
\newtheorem{theorem}{Theorem}

\newtheorem{lemma}{Lemma}

\newtheorem{corollary}{Corollary}

\begin{document}

\title{Caching-at-STARS: the Next Generation Edge Caching}

%\author{
%\IEEEauthorblockN{ Yuanwei~Liu\IEEEauthorrefmark{1}, Zhijin~Qin\IEEEauthorrefmark{1}, Maged Elkashlan\IEEEauthorrefmark{1}, and  Yue~Gao\IEEEauthorrefmark{1}\\} \IEEEauthorblockA{
%\IEEEauthorrefmark{1} Queen Mary University of London, London, UK\\
%%\IEEEauthorrefmark{2} Lancaster University, Lancaster, UK\\
% } }

\author{

% Zhaoming\ Hu, Ruikang\ Zhong, Chao\ Fang, and Yuanwei\ Liu

\normalsize{Zhaoming~Hu,~\IEEEmembership{\normalsize Graduate~Student~Member,~IEEE,}
        Ruikang~Zhong,~\IEEEmembership{\normalsize Graduate Student \\ Member,~IEEE,}
        Chao~Fang,~\IEEEmembership{\normalsize Senior~Member,~IEEE,}
        and Yuanwei~Liu,~\IEEEmembership{\normalsize Senior~Member,~IEEE}}

\thanks{Zhaomin Hu and Chao Fang are with the Faculty of Information Technology, Beijing University of Technology, Beijing, P.R. China, and Purple Mountain Laboratories, Nanjing, P.R. China (email: huzhaoming@emails.bjut.edu.cn; fangchao@bjut.edu.cn).

Ruikang~Zhong and Yuanwei~Liu is with the school of School of Electronic Engineering and Computer Science, Queen Mary University of London, London E1 4NS, U.K. (e-mail: r.zhong@qmul.ac.uk; yuanwei.liu@qmul.ac.uk).
}}

\maketitle
\vspace{-0.5cm}
\begin{abstract}
A simultaneously transmitting and reflecting surface (STARS) enabled edge caching system is proposed for reducing backhaul traffic and ensuring the quality of service. A novel Caching-at-STARS structure, where a dedicated smart controller and cache memory are installed at the STARS, is proposed to satisfy user demands with fewer hops and desired channel conditions. Then, a joint caching replacement and information-centric hybrid beamforming optimization problem is formulated for minimizing the network power consumption. As long-term decision processes, the optimization problems based on independent and coupled phase-shift models of Caching-at-STARS contain both continuous and discrete decision variables, and are suitable for solving with deep reinforcement learning (DRL) algorithm. For the independent phase-shift Caching-at-STARS model, we develop a frequency-aware based twin delayed deep deterministic policy gradient (FA-TD3) algorithm that leverages user historical request information to serialize high-dimensional caching replacement decision variables. For the coupled phase-shift Caching-at-STARS model, we conceive a cooperative TD3 \& deep-Q network (TD3-DQN) algorithm comprised of FA-TD3 and DQN agents to decide on continuous and discrete variables respectively by observing the network external and internal environment. The numerical results demonstrate that: 1) The Caching-at-STARS-enabled edge caching system has advantages over traditional edge caching, especially in scenarios where Zipf skewness factors or cache capacity is large; 2) Caching-at-STARS outperforms the RIS-assisted edge caching systems; 3) The proposed FA-TD3 and cooperative TD3-DQN algorithms are superior in reducing network power consumption than conventional TD3.
\end{abstract}
%\vspace{-1cm}
\begin{keywords}
%\vspace{-0.5cm}
Beamforming, caching replacement, deep reinforcement learning (DRL), edge caching, simultaneously transmitting and reflecting surface (STARS).
\end{keywords}
%\vspace{-0.8cm}
\section{Introduction}

% 流量增多边缘缓存的必要性。
Nowadays, the exponential growth of global mobile network traffic is driven by the explosion of information-centric communications, which encompasses emerging multimedia services such as augmented reality (AR), virtual reality (VR), and metaverse. According to Ericsson mobile report \cite{Eri2023}, global mobile network traffic is predicted to increase fourfold by 2028 compared to 2022, reaching 325 EB per month. Specially, the proportion of 5G traffic in mobile data traffic is expected to be around 69 percent by the end of 2028. The surge in traffic and innovative network technology have prompted network operators to explore new techniques to advance quality of service (QoS) and alleviate backhaul congestion. %https://www.ericsson.com/en/reports-and-papers/mobility-report/dataforecasts/mobile-traffic-forecast

% 边缘缓存介绍以及目前实现边缘缓存在物理层方面的挑战（1、网络边缘实际交付过程中间无线链路不完善，交付率较低且用户移动；2、缓存策略和无线通信的耦合）

%The conventional content caching technologies, which is connection-centric, entails the pre-storing of frequently accessed content in the servers, such as content delivery network (CDN) servers, to reduce content transmission redundancy \cite{Sung2016TMM}.
%With the continuous advancement of wireless communication and emerging future network technologies represented by information-centric networking (ICN) \cite{Chao2015COMST} and software-defined networking (SDN) \cite{Herrera2022JIOT}, \textbf{edge caching} has emerged as a promising technology for wireless networks \cite{Jedari2021COMST}.
Edge caching has emerged as a promising technology for wireless networks with the continuous advancement of wireless communication and emerging future network architectures represented by information-centric networking~(ICN)~ \cite{Chao2015COMST} and software-defined networking~(SDN)~\cite{Herrera2022JIOT}. It deploys caches at network edge nodes, such as base stations (BSs) or roadside units (RSU), which are closer to the users' physical location, to reduce content transmission redundancy \cite{Jedari2021COMST}. Nevertheless, the quality of wireless links at the network edge are often flawed due to factors such as the inherent service capabilities of transceivers, which can adversely affect the user's quality of experience (QoE) and the design of caching strategies \cite{Zhang2018MWC}. For instance, in scenarios such as at the cellular coverage edge or in environments with numerous obstacles, the signal transmission rate received by the user is likely to be sluggish, and there is a higher likelihood of decoding failure. Therefore, the realization of edge caching in wireless networks have to face the coupling problem of caching strategy and wireless communication, and it is crucial to enhance the performance of edge caching from the perspective of wireless communication.

% RIS是一个无线链路交付解决方案，RIS只服务单面用户，RIS 基础上提出STARS。STARS自身可以进行边缘缓存。

Reconfigurable intelligent surface (RIS) is a novel technique that has the potential to enhance the coverage, signal quality, and energy efficiency of wireless networks \cite{Liu2021COMST}. Typically, RIS is comprised of massive programmable elements that can adapt the amplitude and phase-shift of electromagnetic waves as necessary to facilitate improved signal-to-interference-plus-noise ratio~(SINR). In practical scenarios, the reflecting-only RIS can provide signal enhancement for users on the same side as the BS by reflecting and focusing the signal towards them. However, for users on the opposite side, the signals are weaker due to the attenuation and scattering~\cite{Zhong2023JSTSP}. To address this challenge, a simultaneous transmitting and reflecting surface~(STARS) is proposed to achieve $360^\circ$ global coverage around the RIS panel \cite{Liu2021MWC,Zhong2022JSACNOMA}. The STARS is capable of simultaneously transmitting and reflecting signals, providing a high degree of freedom (DoF) that enables efficient utilization of spectrum resources and improvement of data transmission rates. Moreover, STARS is typically deployed on the user side, expanding the edge cache on it can effectively reduce the distance of data transmission and enhance security \cite{Mu2022JSAC}.

To integrate STARS into the edge caching system, two possible system models can be considered: 1) Employing STARS to assist the wireless communication module in the edge caching system; 2) Utilizing STARS for both content caching and assisting wireless communication. The second method provides a larger distributed caching space in the network edge and reduces hop counts of user fetched requests. With the deployment of the cache on STARS, user requests can be fetched from more diverse edge nodes, which makes caching replacement decisions directly affect the control of STARS amplitude and phase-shift. Therefore, it is necessary to design beamforming based on the caching information of edge nodes, which is referred to as \textbf{information-centric hybrid beamforming}. Implementing information-centric hybrid beamforming on edge caching system can adaptively adjust beamforming according to network status, avoiding the energy consumption caused by a single beamforming paradigm. Joint optimization of caching replacement and information-centric hybrid beamforming is a long-term decision-making process, which requires real-time observation of network state information and user state information for dynamic decision-making. Fortunately, the rapid development of deep reinforcement learning~(DRL) provides the possibility of dynamic decision-making in this complex environment~\cite{Chen2021COMST}. Hence, DRL supporting joint optimization of caching replacement and information-centric hybrid beamforming are envisioned to improve the content distribution efficiency.
\vspace{-0.2cm}
%The joint implementation of caching and hybrid beamforming on STARS involves optimizing caching, communication, and computing resources simultaneously.
\subsection{The State of the Art}

%Video Caching, Analytics, and Delivery at the Wireless Edge: A Survey and Future Directions

Edge caching, as a research hotspot in recent years, can effectively reduce network throughput and delay by storing popular content in network edge nodes that are closer to users during off-peak hours. Furthermore, wireless communication technologies have also changed with the blessing of edge caching \cite{Wen2018TCOMM,Cao2020LWC,He2019LWC,Xiao2022JSAC,Raptis2020JSAC}. %Hybrid Content Caching in 5G Wireless Networks: Cloud Versus Edge Caching.
Specifically, Cao \emph{et al.} \cite{Cao2020LWC} focused on caching replacement and multi-antenna multiplexing for multiple-input multiple-output (MIMO) fog radio access networks (F-RANs) with small caching capacity.
He \emph{et al.} \cite{He2019LWC} investigated information-centric coordinated fronthaul data assignment and multicast beamforming for wireless networks.
Xiao \emph{et al.} \cite{Xiao2022JSAC} presented a cache enabled two-tier non-orthogonal multiple access (NOMA)-based BS-multicast group matching mechanism to solve the collaboration challenge during the edge delivery process.
P. Raptis \emph{et al.} \cite{Raptis2020JSAC} studied distributed data access in multi-hop wireless industrial edge networks to increase energy efficiency.
Aforementioned studies have effectively addressed wireless transmission issues in various practical edge caching network scenarios, but they may not be applicable to scenarios with significant wireless signal fading.

To address this issue, some researchers have investigated edge caching in scenarios with significant signal fading. Chhangte \emph{et al.} \cite{Chhangte2021TNSM} regarded WiFi router as a edge caching node and proposed a Wi-Cache protocol to decrease network throughput.
Zhang \emph{et al.} \cite{Zhang2022TCOMM} designed a cooperative caching architecture of unmanned aerial vehicle (UAV) and user terminal to facilitate UAV-assisted end-to-end communication.
In addition, for the RIS-assisted edge caching, Chen \emph{et al.} \cite{Chen2021TWC} were committed to using RIS to assist edge caching system so that enhance the QoS at the coverage boundary of wireless networks. Mei \emph{et al.} \cite{Mei2021LWC} tried to study the joint optimization of UAV trajectory, task offloading and cache with the phase-shift design of the RIS in mobile edge computing (MEC). However, in this study, RIS is only utilized to assist the wireless communication network and is not tightly integrated with the edge caching system. Additionally, these researches are constrained by the inherent service characteristics of RIS.

The swift advancement of artificial intelligence (AI) and machine learning (ML) has the potential to enhance network performance through AI empowered edge caching \cite{Wen2018TCOMM,Bura2022TNET,Li2023TWC,Fu2021TWC}.
Fu \emph{et al.} \cite{Fu2022JSAC} explore the use of recommendation algorithm to predict user behavior and assist cache-enabled NOMA networks.
Yang \emph{et al.} \cite{Yang2022TWC} adopted long short-term memory (LSTM) to forecast events popularity and proposed a NOMA-based content transmission scheme.
Jiang \emph{et al.}~\cite{Jiang2020TWC} proposed a deep learning-based popularity trend classification and user location prediction model to guide edge caching in F-RANs.

Specially, DRL, an ML algorithm that excels in long-term decision-making, is widely used in caching replacement \cite{Chao2022JIOT,Yang2020TCOMM,Wu2021TWC} and intelligent beamforming control of RIS \cite{Wang2020COMST,Huang2020JSAC,Yang2021TWC,Zhong2022TWC}.
In DRL-based caching replacement, Li \emph{et al.} \cite{Li2023TITS} employed dueling deep Q network (Dueling-DQN) to optimize the proposed federated edge cooperative caching scheme.
Tian \emph{et al.} \cite{Tian2023TNSM} presented a DRL-cache admission algorithm to optimize cooperative caching for mobile edge networks.
In DRL-based intelligent beamforming control of RIS, Samir \emph{et al.} \cite{Samir2021TVT} investigated the passive beamforming at the RIS through the proximal policy optimization (PPO) algorithm.
Zhong \emph{et al.}~\cite{Zhong2022JSACRL} presented two improved deep deterministic policy gradient (DDPG) algorithms to control the beamforming of the STARS in multi-user downlink multiple-input single-output (MISO) communication system.
However, the adoption of DRL for joint optimization of edge caching replacement and information-centric hybrid beamforming control still remains a challenging task.
\vspace{-1cm}
\subsection{Motivation and Challenges}

Compared with the traditional edge caching, cache-empowered STARSs, namely \textit{Caching-at-STARSs}, are utilized to assist edge caching system not only by intelligently controlling wireless signals to enhance coverage and signal quality of wireless communication, but also by sinking the cache capacity to the user side to fulfill user demands with fewer hops and desirable channel conditions. However, designing a joint decision for caching replacement and hybrid beamforming on Caching-at-STARS still poses a challenge. Firstly, the hybrid beamforming on STARS is influenced by the cache status of network nodes in the edge caching system, necessitating the design of information-centric hybrid beamforming for Caching-at-STARS. In the conventional edge caching system, users fetch the requested content from the BS or remote server based on the cache state at the BS. However, when STARS is empowered with caching capability, user requests may be fetched at STARS, BS, and remote servers, resulting in the hybrid beamforming is tightly coupled to the cache status of both STARS and BS. Secondly, most of the existing studies have assumed that reflection and transmission phase-shift control of STARS is independent. However, in reality, due to the electromagnetic characteristics of STARS elements, the electric and magnetic impedances cannot be arbitrarily set, which results in the need for coupling control of the transmission and reflection phase-shift (T\&R phase-shift) of STARS. Therefore, we will study the effects of independent and coupled T\&R phase-shifts on the Caching-at-STARS, respectively. Thirdly, in the edge caching system with Caching-at-STARS, caching replacement and coupled STARS phase-shifts control are discrete variables, while the remaining decision variables are continuous. Traditional DRL cannot simultaneously optimize continuous and discrete variables. Therefore, we adopt two ideas: 1) To convert discrete variables into continuous variables; 2) To design a new DRL algorithm that can simultaneously optimize continuous and discrete variables to jointly optimize caching replacement and information-centric hybrid beamforming.
%1、直接在距离用户更近的STARS上部署缓存可以更快地响应用户请求，减少跳数，STARS的缓存情况也会影响无线信道

%2、STARS可以服务两侧的用户而不是单边用户

%3、independent 和 coupled STARS同时考虑

%4、DRL同时缓存和beamfroming的长期决策
\vspace{-0.2cm}
\subsection{Contributions}

The primary contributions in this paper are detailed as follows:
\begin{itemize}
\item[$\bullet$] We propose a novel STARS-enabled edge caching system with a Caching-at-STARS structure, which enables caching on STARS and can intelligently control wireless signals. Based on this design, we formulate the problem of minimizing network power consumption in edge caching enabled by the Caching-at-STARS to jointly optimize edge node caching replacement and information-centric hybrid beamforming. In addition, the independent and coupled T\&R phase-shifts models for Caching-at-STARS are considered separately in the optimization problem.
\item[$\bullet$] For the independent T\&R phase-shift Caching-at-STARS, we propose a frequency-aware based twin delayed deep deterministic policy gradient (FA-TD3) algorithm. Due to the high-dimensional nature of content caching decision variables, we design a frequency-aware dynamic continuous strategy, which automatically adjust the decision scale of in-network content in the caching action space according to user historical request information. Therefore, the entire action space in the TD3 algorithm is transformed into a continuous space for joint optimization of caching replacement and wireless signal control.
\item[$\bullet$] For coupled T\&R phase-shift Caching-at-STARS, we conceive a cooperative TD3 \& deep-Q network (TD3-DQN) algorithm. Compared with the independent T\&R phase-shift Caching-at-STARS model, the coupled model adds a constraint on the transmission phase-shift. For these additional discrete variables, we constructed FA-TD3 and DQN co-agents to solve the problem, in which the entire environment is divided into the external real environment and the internal virtual environment. The FA-TD3 agent is used to optimize continuous variables in the external environment, and the DQN agent makes decisions on discrete variables in the internal environment by perceiving the action and state information feedback from the external environment.
\item[$\bullet$] Numerical results reveal that 1) The proposed Caching-at-STARS outperforms both STARS-aided edge caching and STARS without edge caching, particularly in scenarios with large Zipf skewness factor and cache capacity, since users have a greater probability of fetching the request from STARS closer to them; 2) STARS assisted edge caching achieves superior performance than RIS assisted edge caching; 3) For independent T\&R phase-shift Caching-at-STARS, FA-TD3 algorithm outperforms than conventional TD3 algorithm; 4) For coupled T\&R phase-shift Caching-at-STARS, cooperative TD3-DQN algorithm outperforms the conventional TD3 algorithm.
\end{itemize}
\vspace{-0.7cm}
\subsection{Organizations}

This paper is organized as follows. Section II illustrates the model of narrowband Caching-at-STARS-enabled downlink edge caching system, including the caching model, independent and coupled T\&R phase-shift STARS model, signal model, and problem formulation. Section III proposes a FA-TD3 algorithm for the independent T\&R phase-shift Caching-at-STARS, while Section IV develops a cooperative TD3-DQN algorithm for the coupled T\&R phase-shift Caching-at-STARS. Section V discusses the numerical results. Finally, we conclude this study in Section VI.

\section{System Model} \label{system_model}

\begin{figure}[tp]
\centering
\includegraphics[scale=0.4]{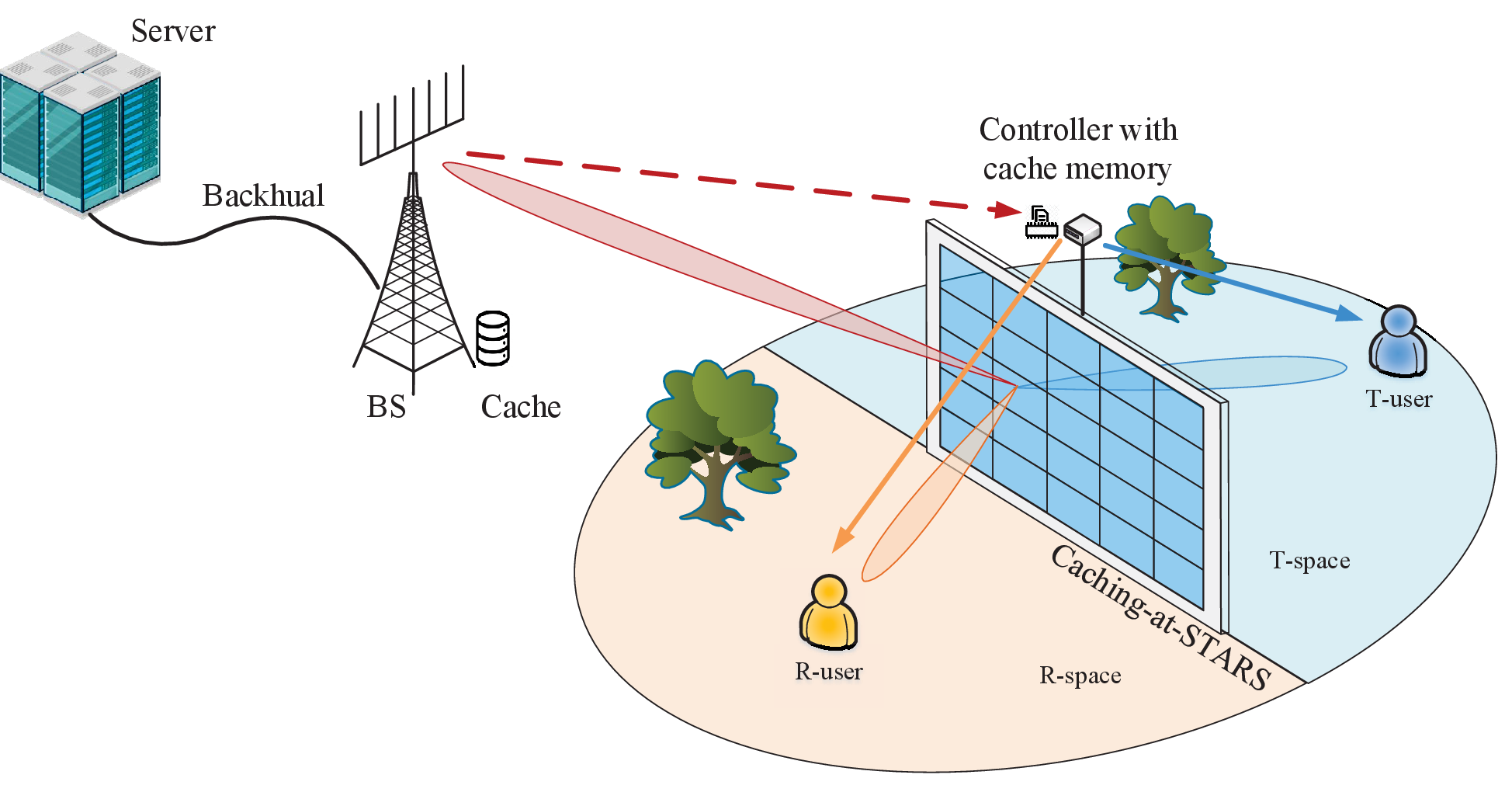}
\caption{System model of Caching-at-STARS-enabled downlink communication system.}
\label{fig:system model}
\vspace{-1cm}
\end{figure}

As shown in Fig.~\ref{fig:system model}, we consider a narrowband Caching-at-STARS-enabled downlink edge caching system, where an $M$-antenna BS communicates with single-antenna users. The Caching-at-STARS, which includes $N$ passive transmission-reflection (T-R) elements, is utilized to enhance wireless communication. Meanwhile, these T-R elements connected to a smart controller with single antenna using a single wire link. Besides adjusting the STARS transmission and reflection, the smart controller equipped with cache memory can also acts as a potential decode-and-forward relay and actively transmits caching content to the target user. Thus, we refer to this new STARS with smart controller and cache memory as the Caching-at-STARS. For a given STARS, the whole space is divided into two half spaces by it, namely the transmission space (T-space) and reflection space (R-space). Likewise, the users that located in the T-space and R-space are referred to as T-users and R-users.
%The antenna of the controller is oriented towards the R-space to facilitate region partitioning consistent with the overall model when STARS transmits cached content directly to the user.
The directly links between BS and users are assumed to be blocked. Moreover, we consider a coherent time slot of the length $L$, during which the communication channels and user request content remain approximately constant.

\vspace{-0.5cm}
\subsection{Caching Model}

We assume there exists a catalog $\mathcal{F}=\left\{1,2,...,F\right\}$ of contents (such as files or file blocks) in the network, where each content has the normalized size of 1. The in-network contents follows Zipf distribution with the skewness factor $\alpha$. All content in catalog $\mathcal{F}$ sequenced according to their popularity, with the most popular being ranked 1-st and the least popular being ranked $F$-th. Therefore, the probability of the $f$-th content is requested by user is $p_f = \frac{f^{-\alpha}}{\textstyle \sum_{\chi=1}^F \chi^{-\alpha}}$. For simplicity, we assume that the distribution of requests for T-user $k_t$ and R-user $k_r$ is the same, i.e. $p_f^{k_t} = p_f^{k_r} \triangleq p_f$. Both BS and Caching-at-STARS are deployed storage capabilities with a maximum caching capacity of $C_b$ and $C_c$ ($C_c<C_b<F$), respectively.
%The cache state vector $\mathbf{c}_{t}^i = \left\{c_{1,t}^i,...,c_{l,t}^i,...,c_{L_i,t}^i\right\}$, where $i \in \mathcal{I}=\left\{b,c\right\}$, and $c_{l,t}^i \in \mathcal{F}$.
The cache state vector $\mathbf{c}_{t}^i = \left\{c_{1,t}^i,...,c_{f,t}^i,...,c_{F,t}^i\right\}$, where $i \in \mathcal{I}=\left\{b,c\right\}$, and $c_{f,t}^i \in \left\{0,1\right\}$. The $c_{f,t}^i = 1$ represents the network node $i$ caches content $f$ at time slot $t$, otherwise, $c_{f,t}^i = 0$ denotes that the node $i$ does not caches. Due to the limited caching capacity of the BS and Caching-at-STARS, $\textstyle \sum_{f\in\mathcal{F}} c_{f,t}^i \leq C_i$ need to be satisfied. Edge caching consists of two main phases: content pushing and content delivering. Content pushing involves pushing cached content from a remote server to a BS or Caching-at-STARS and can be considered as a pre-fetching process. Content delivering, on the other hand, refers to the process of sending content from the BS or Caching-at-STARS to the user over the wireless network in response to a user request. Since the location of the BS and Caching-at-STARS is fixed and the BS and remote server are wired together, the push power consumption per unit of content replacement is simplified to $P_u$.
Due to this paper focuses on wireless communication scenarios, when neither the BS nor STARS caches user requests, they will be transmitted through the wired link to the remote server. The cost of a single content fetch from remote server to BS via backhaul will be simplified to $P_{bh}$.

%The frequency of content pushing is significantly lower than the frequency of content delivering, especially during peak hours, and this model does not consider scenarios containing multiple Caching-at-Stars. Therefore, this paper ignores the power consumption of content pushing\footnote{As only a single Caching-at-STARS is included in this model, there is no need to consider interference during content pushing phase. The location of the BS and Caching-at-STARS is kept constant resulting in the power consumption of content push being only related to the amount of replacement cached content. Furthermore, the frequency of content pushing is significantly lower than the frequency of content delivering. To emphasize the control of signal transmission, the power consumption of content pushing phase is ignored in this paper.} and focuses on power allocation and cache optimisation during the content distribution phase.

%\begin{align}\label{Cachestate}
%X_j = \left\{ \begin{array}{rcl}
%
%\end{align}
\vspace{-0.2cm}
\subsection{STARS Model}

We adopt an energy splitting model for supporting simultaneous transmission and reflection of the Caching-at-STARS. In addition, the passive T-R elements of STARS do not actively transmit signals. Signals transmitted by the BS and Caching-at-STARS controller are passively beamforming through STARS for signal conditioning. The incident signal is splitting by Caching-at-STARS into transmission signal and reflection signal according to its recipient users. Denote $\mathbf{\Theta}_{\mathcal{T},t} \in \mathbb{C}^{N \times N}$ and $\mathbf{\Theta}_{\mathcal{R},t} \in \mathbb{C}^{N \times N}$ as the matrices of the transmission coefficients (TCs) and reflection coefficients (RCs) at each time slot $t$, respectively, which can be modeled as
%\begin{align}\label{TCs}
%\mathbf{\Theta}_{\mathcal{T},t} = diag(\beta_{\mathcal{T},1,t} e ^{j\theta_{\mathcal{T},1,t}},\beta_{\mathcal{T},2,t} e ^{j\theta_{\mathcal{T},2,t}},...,\beta_{\mathcal{T},N,t} e ^{j\theta_{\mathcal{T},N,t}}),
%\end{align}
%\begin{align}\label{TCs}
%\mathbf{\Theta}_{\mathcal{R},t} = diag(\beta_{\mathcal{R},1,t} e ^{j\theta_{\mathcal{R},1,t}},\beta_{\mathcal{R},2,t} e ^{j\theta_{\mathcal{R},2,t}},...,\beta_{\mathcal{R},N,t} e ^{j\theta_{\mathcal{R},N,t}}),
%\end{align}
\begin{align}\label{TCs}
\mathbf{\Theta}_{q,t} = diag(\beta_{q,1,t} e ^{j\theta_{q,1,t}},\beta_{q,2,t} e ^{j\theta_{q,2,t}},...,\beta_{q,N,t} e ^{j\theta_{q,N,t}}),
\end{align}
where $q \in \mathcal{Q} = \left\{\mathcal{T}, \mathcal{R}\right\}$, $\beta_{q,n,t} \in [0,1]$, and $\theta_{q,n,t}\in [0,2\pi]$ denote the amplitude and phase-shift response of the $n$-th element. Moreover, according to the \textit{law of energy conservation}, the amplitudes coefficient should be satisfied:
\begin{align}\label{TCs}
\beta_{q,n,t}^2 + \beta_{\overline{q},n,t}^2 = 1, n = 1,2,\cdots,N.
\end{align}

Generally, the model in which the phase-shifts of TCs and TRs can be adjusted independently is termed as \textit{independent T\&R phase-shift models}. However, this control requires the elements of Caching-at-STARS to be active, which is likely to lead to higher manufacturing costs. As such, the low-cost and lossless elements has been investigated, where the electric and magnetic impedances of each element should be purely imaginary. Therefore, the following conditions should be satisfied for TCs and RCs:
\begin{align}\label{TCs}
cos(\theta_{q,n,t}-\theta_{\overline{q},n,t})=0, n = 1,2,\cdots,N.
\end{align}
In the above constraints, the model is referred as the \textit{coupled T\&R phase-shift models}. To fully investigate the effect of STARS on the system, the independent and coupled T\&R phase-shift Caching-at-STARS models are considered in this paper.
\vspace{-0.2cm}
\subsection{Signal Model}

\begin{figure}[tp]
\centering
\includegraphics[scale=0.8]{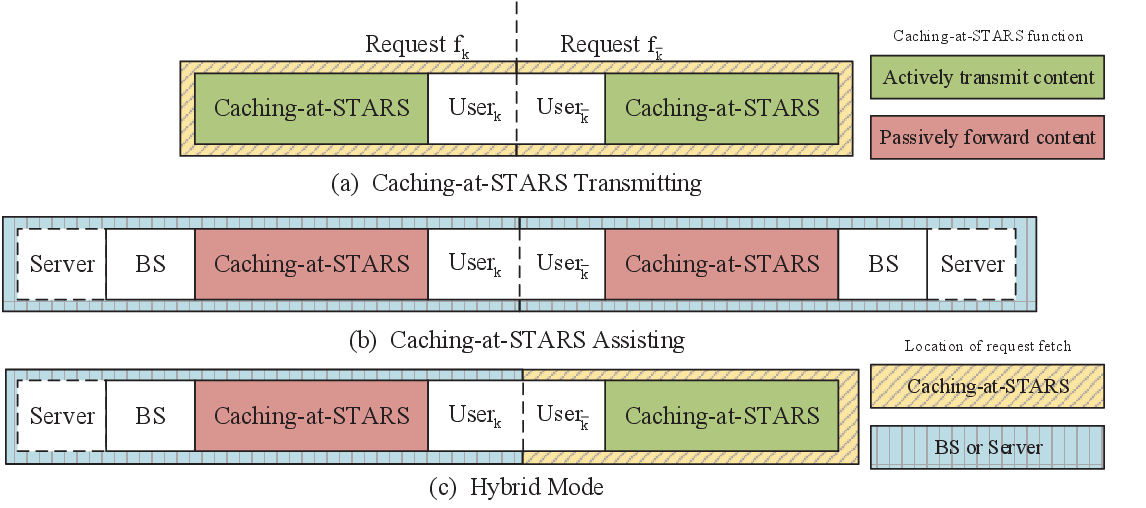}
\caption{The hybrid beamforming protocols for the Caching-at-STARS aided wireless communication.}
\label{fig:STARS_mode}
\end{figure}

We consider multiple channels in Caching-at-STARS aided wireless communication system, including the BS to Caching-at-STARS passive elements channel $\mathbf{G}_{b,t} \in \mathbb{C}^{N \times M}$,
%Caching-at-STARS controller to passive elements channel $\mathbf{g}_{c,t} \in \mathbb{C}^{N \times 1}$.
the passive T and R channel of T-user $k_t$ or R-user $k_r$ at time slot $t$ are $\mathbf{h}_{c,k_t,t} \in \mathbb{C}^{N \times 1}$ and $\mathbf{h}_{c,k_r,t}\in \mathbb{C}^{N \times 1}$,
and the direct channels from the single antenna of the Caching-at-STARS to the T-user $k_t$ or R-user $k_r$ at time slot $t$ are $h_{c,k_t,t}^d \in \mathbb{C}^{1 \times 1}$ and $h_{c,k_r,t}^d \in \mathbb{C}^{1 \times 1}$. The information-bearing symbol and the active beamforming vectors for content $f$ sent to user $k$ at the BS are denoted by $s_{f,k,t}^b$, $\mathbf{P}_{k,t}^b$, where $k \in \mathcal{K} = \left\{k_t, k_r\right\}$. Moreover, the information-bearing symbol and the active transmission power are denoted by $s_{f,k,t}^c$, $P_{k,t}^c$.
Since all content is the same size by default, it is assumed that the beamforming vector and direct transmit power are independent of the content being transmitted.
%In the following, we will illustrate the signal models for the CT, CA and HM modes.
According to the caching status of Caching-at-STARS and BS, we consider three hybrid beamforming protocols for content delivering phases in the Caching-at-STARS aided wireless communication, namely Caching-at-STARS transmitting (CT), Caching-at-STARS assisting (CA), and Hybrid mode (HM), refers to Fig.~\ref{fig:STARS_mode}.

1) \textbf{Caching-at-STARS transmitting}: For CT, the Caching-at-STARS controller transmits the content to the user through the T-R elements if the user request content can be retrieved in the STARS' cache.

2) \textbf{Caching-at-STARS assisting}: For CA, the BS delivers the content to users with Caching-at-STARS assistance when STARS does not cache the corresponding content and the BS or server does.

3) \textbf{Hybrid mode}: For HM, the Caching-at-STARS performs both the function of assisting the BS in delivering content, and its controller sends content to the user when the user's request needs to be fetched on the STARS or BS respectively.

According to the Zipf distribution, popular content has a higher probability of being requested by different users in the network. Therefore, within each mode we need to distinguish between scenarios where the user requests the same content and different content.

1) CT: The Caching-at-STARS caches user request contents, i.e. $c_{f,t}^c = c_{\overline{f},t}^c = 1$, $f$ and $\overline{f}$ are the contents requested by the two users at time slot $t$, respectively. In this mode, we can regard the system as a multi-users SISO system. Caching-at-STARS controller transmits the content to the user through the direct channels from controller to users. The received signal of user $k$ at time slot $t$ is given by
\begin{align}\label{CT_t}
y_{k,t}^{\text{CT}} = h_{c,k,t}^d (\sqrt{P_{k,t}} s_{f,k,t}^c + \sqrt{P_{\overline{k},t}} s_{\overline{f},\overline{k},t}^c)+n_k,
\end{align}
where $\mathbb{E}\left[\left|s_{f,k,t}^c\right|^2\right]=1$, and $n_k \in \mathcal{C} \mathcal{N}(0,\sigma_k^2)$ denotes the Gaussian noise. Meanwhile, $f = f_r$, $\overline{k} = k_t$, $\overline{f} = f_t$, if $k = k_r$; and $f = f_t$, $\overline{k} = k_r$, $\overline{f} = f_r$, otherwise.
If T-user and R-user request the different content, the achievable communication rate of user $k$ at time slot $t$ is given by
\begin{align}\label{RCT_rD}
R_{k,t}^{\text{CT}} = Blog_2 \left(1+\frac{|h_{c,k,t}^d|^2 P_{k,t}^c}{|h_{c,k,t}^d |^2 P_{\overline{k},t}^c +\sigma^2}\right),
\end{align}
where $B$ is bandwidth, $\sigma^2$ represents the noise power. Conversely, when T-user and R-user request the same content, their corresponding rata are
\begin{align}\label{RCT_tS}
R_{k,t}^{\text{CT}} = Blog_2 \left(1+\frac{|h_{c,k,t}^d|^2 P_{k,t}^c}{\sigma^2}\right).
\end{align}

Therefore, the wireless transmission power of the system in CT at time slot $t$ can be expressed as
\begin{align}\label{RCT_rS}
P_{w,t} = P_{k,t}^c + P_{\overline{k},t}^c, \qquad if\ c_{f,t}^c = c_{\overline{f},t}^c = 1.
\end{align}

2) CA: The Caching-at-STARS doesn't cache user request, which need to be fetched at the BS or remote server, i.e. $c_{f,t}^c = c_{\overline{f},t}^c = 0$. The Caching-at-STARS plays a role in assisting wireless transmission, the user received signal is given by
%the received signal of T-user $k_t$ and R-user $k_r$ at slot $t$ is given by
%\begin{align}\label{CA_r}
%y_{k_t,t}^{CA} = \mathbf{h}_{c,k_t,t} \mathbf{\Theta}_{\mathcal{T},t} \mathbf{G}_b \mathbf{w}_{b,k_t,t} s_{b,k_t,t}+n_0,
%\end{align}
%\begin{align}\label{CA_r}
%y_{k_r,t}^{CA} = \mathbf{h}_{c,k_r,t} \mathbf{\Theta}_{\mathcal{R},t} \mathbf{G}_b \mathbf{w}_{b,k_r,t} s_{b,k_r,t}+n_0.
%\end{align}
\begin{align}\label{CA}
y_{k,t}^{\text{CA}} = \mathbf{h}_{c,k,t}^H \mathbf{\Theta}_{q,t} \mathbf{G}_{b,t} (\mathbf{P}_{k,t}^b s_{f,k,t}^b+\mathbf{P}_{\overline{k},t}^b s_{\overline{f},\overline{k},t}^b)+n_k.
\end{align}

When T-user and R-user request the different content, the achievable communication rate of user $k$ at time slot $t$ is given by
%\begin{align}\label{RCA_tD}
%R_{k_t,t}^{CA} = Blog_2 \left(1+\frac{|\mathbf{h}_{c,k_t,t} \mathbf{\Theta}_{\mathcal{T},t} \mathbf{G}_b \mathbf{w}_{b,k_t,t}|^2}{|\mathbf{h}_{c,k_t,t} \mathbf{\Theta}_{\mathcal{T},t} \mathbf{G}_b \mathbf{w}_{b,k_r,t}|^2+\sigma^2}\right),
%\end{align}
%\begin{align}\label{RCA_rD}
%R_{k_r,t}^{CA} = Blog_2 \left(1+\frac{|\mathbf{h}_{c,k_r,t} \mathbf{\Theta}_{\mathcal{R},t} \mathbf{G}_b \mathbf{w}_{b,k_r,t}|^2}{|\mathbf{h}_{c,k_r,t} \mathbf{\Theta}_{\mathcal{R},t} \mathbf{G}_b \mathbf{w}_{b,k_t,t}|^2+\sigma^2}\right).
%\end{align}
\begin{align}\label{RCA_rD}
R_{k,t}^{\text{CA}} = Blog_2 \left(1+\frac{|\mathbf{h}_{c,k,t}^H \mathbf{\Theta}_{q,t} \mathbf{G}_{b,t} \mathbf{P}_{k,t}^b|^2}{|\mathbf{h}_{c,k,t}^H \mathbf{\Theta}_{q,t} \mathbf{G}_{b,t} \mathbf{P}_{\overline{k},t}^b|^2+\sigma^2}\right),
\end{align}
where $\mathcal{J}=\mathcal{T}$, if $k = k_t$; $\mathcal{J}=\mathcal{R}$, otherwise.
When T-user and R-user request the same content, their achievable communication rate are
%\begin{align}\label{RCA_tS}
%R_{k_t,t}^{CA} = Blog_2 \left(1+\frac{|\mathbf{h}_{c,k_t,t} \mathbf{\Theta}_{\mathcal{T},t} \mathbf{G}_b \mathbf{w}_{b,k_t,t}|^2}{\sigma^2}\right),
%\end{align}
%\begin{align}\label{RCA`1111_rS}
%R_{k_r,t}^{CA} = Blog_2 \left(1+\frac{|\mathbf{h}_{c,k_r,t} \mathbf{\Theta}_{\mathcal{R},t} \mathbf{G}_b \mathbf{w}_{b,k_r,t}|^2}{\sigma^2}\right).
%\end{align}
\begin{align}\label{RCA_rS}
R_{k,t}^{\text{CA}} = Blog_2 \left(1+\frac{|\mathbf{h}_{c,k,t}^H \mathbf{\Theta}_{q,t} \mathbf{G}_{b,t} \mathbf{P}_{k,t}^b|^2}{\sigma^2}\right).
\end{align}

Therefore, the wireless transmission power of the system in CA at time slot $t$ can be expressed as
\begin{align}\label{RCT_rS}
P_{w,t} = \parallel \mathbf{P}^b_{{k,t}} \parallel^2 + \parallel \mathbf{P}^b_{\overline{k},t} \parallel^2, \qquad if\ c_{f,t}^c = c_{\overline{f},t}^c = 0.
\end{align}

3) HM: The user request content will be obtained at STARS and the BS or remote server, respectively, i.e. $c_{f,t}^c \oplus c_{\overline{f},t}^c = 1$. The content of user requests is inevitably different since the request content is processed at different network nodes.
In this mode, the received signal of user $k$ fetching content at Caching-at-STARS and user $\overline{k}$ fetching content at BS at time slot $t$ can be model as
\begin{align}\label{HM_t}
y_{k,t}^{\text{HM}} = \underbrace{h_{c,k,t}^d \sqrt{P_{k,t}} s_{f,k,t}^c}_{\text{desired signal}}+\underbrace{\mathbf{h}_{c,k,t}^H \mathbf{\Theta}_{q,t} \mathbf{G}_{b,t} \mathbf{P}_{\overline{k},t}^b s_{\overline{f},\overline{k},t}^b}_{\text{interference}}+n_k,
\end{align}
\begin{align}\label{HM_t}
y_{\overline{k},t}^{\text{HM}} = \underbrace{\mathbf{h}_{c,\overline{k},t}^H \mathbf{\Theta}_{\overline{q},t} \mathbf{G}_{b,t} \mathbf{P}_{\overline{k},t}^b s_{\overline{f},\overline{k},t}^b}_{\text{desired signal}}+\underbrace{h_{c,\overline{k},t}^d \sqrt{P_{k,t}}s_{f,k,t}^c}_{\text{interference}}+ n_{\overline{k}}.
\end{align}
%\begin{align}\label{HM_r}
%y_{k_r,t}^{HM} = \left(\mathbf{h}_{c,k_r,t}^d + \mathbf{h}_{c,k_r,t} \mathbf{\Theta}_{\mathcal{R},t} \mathbf{g}_c \right) \mathbf{w}_{f,k_r,t}^c s_{f,k_r,t}^c+n_0,
%\end{align}

The achievable communication rate of two users is given by
\begin{align}\label{RCT_tD}
R_{k,t}^{\text{HM}} = Blog_2 \left(1+\frac{|h_{c,k,t}^d|^2 P_{k,t}^c}{|\mathbf{h}_{c,k,t}^H \mathbf{\Theta}_{q,t} \mathbf{G}_{b,t} \mathbf{P}_{\overline{k},t}^b|^2+\sigma^2}\right),
\end{align}
\begin{align}\label{RCT_rD}
R_{\overline{k},t}^{\text{HM}} = Blog_2 \left(1+\frac{|\mathbf{h}_{c,\overline{k},t}^H \mathbf{\Theta}_{\overline{q},t} \mathbf{G}_{b,t} \mathbf{P}_{\overline{k},t}^b|^2}{|h_{c,\overline{k},t}^d|^2 P_{k,t}^c+\sigma^2}\right).
\end{align}

Therefore, the wireless transmission power of the system in HM at time slot $t$ can be expressed as
\begin{align}\label{RCT_rS}
P_{w,t} = P_{k,t}^c + \parallel \mathbf{P}^b_{{\overline{k},t}} \parallel^2 , \qquad  if\ c_{f,t}^c \oplus c_{\overline{f},t}^c = 1.
\end{align}

The power consumption of the whole system consists of backhaul power consumption, wireless content delivering power consumption, and wireless content pushing power, which can be formulated as
\begin{align}\label{RCT_rS}
P_{s,t} = P_{w,t} + \lambda_{r,t} P_{bh} + \lambda_{u,t} P_{u},
\end{align}
where $\lambda_{r,t} = \sum_{k \in \mathcal{K}} \overline{c_{f_{k},t}^b \vee c_{f_{k},t}^c}$ represents the number of requests processed at the remote server, $\lambda_{u,t} = \sum_{i \in \mathcal{I}} |\mathbf{c}_{t}^i - \mathbf{c}_{t-1}^i| $ represents the number of replaced contents in the cache.
\vspace{-0.2cm}

\subsection{Problem formulation}

We aim to optimize the active beamforming vector $\mathbf{P}^b$, active Caching-at-STARS transmission power $P^c$, STARS' T-R coefficient matrix $\mathbf{\Theta}$, and content caching vector $\mathbf{c}$ to minimize the network power consumption, which consists of the backhaul power consumption and the wireless transmission power. To this end, we formulate the optimization problem in independent T\&R phase-shift Caching-at-STARS model as \textbf{Problem 1}:
\begin{subequations}
\begin{align}\label{OPP1}
%&\min_{\mathbf{c}_{b,t},\mathbf{c}_{c,t},\mathbf{w}_{f,k_t,t}, \mathbf{w}_{f,k_r,t}, \mathbf{\Theta}_{\mathcal{T},t},\mathbf{\Theta}_{\mathcal{R},t}} \sum_{t=1}^{T} \parallel \mathbf{w}^2_{\mathbf{f},t}\parallel + \lambda_t w_b,\\
%\mathcal{P}: & \min_{\left\{\mathbf{P}^b_{{k,t}}, P_{k,t}^c, \mathbf{\Theta}_{q,t}, \mathbf{c}_{t}^i\right\}} \sum_{t=1}^{T} \left( P_{w,t} + \lambda_{r,t} P_{bh} + \lambda_{u,t} P_{u}\right),\\
\text{P1}:  \min_{\mathbf{P}^b, P^c, \mathbf{\Theta}, \mathbf{c}} \quad & \sum_{t=1}^{T} P_{s,t},\\
\textrm{s.t.} \; \ \
%& -\pi \leq \theta_{\mathcal{R},n,t} \le \pi, \forall n, \forall t,\label{OPPC}\\
& 0 \leq \theta_{q,n,t} \le 2\pi, \forall q, \forall n, \forall t,\label{OPP1B}\\
&\beta_{q,n,t}^2+\beta_{\overline{q},n,t}^2=1,0 \leq \beta_{q,n,t}\leq 1, \forall q, \forall n, \forall t,\label{OPP1E}\\
& R_{k,t}^{\text{X}} \geq R_\text{QoS}, \forall k, \forall t, \label{OPP1D}\\
%& \beta_{n,t} \sqrt{1- \beta_{n,t}^2} \cos(\theta_{\mathcal{R},n,t} - \theta_{\mathcal{T},n,t}) = 0, \label{OPPF}\\
& \sum_{f\in\mathcal{F}} c_{f,t}^i \leq C_i, \forall i, \forall t, \label{OPP1C}\\
& P_{t}^i \leq P_{\text{max}}^i, \forall i, \forall t, \label{OPP1G}
\end{align}
\end{subequations}
%where $\lambda_{r,t} = \sum_{k \in \mathcal{K}} \overline{c_{f_{k},t}^b \vee c_{f_{k},t}^c}$ represents the number of requests processed at the remote server, $\lambda_{u,t} = \sum_{i \in \mathcal{I}} |\mathbf{c}_{t}^i - \mathbf{c}_{t-1}^i| $ represents the number of replaced contents in the cache, and $\text{X} \in \left\{\text{CT}, \text{CA}, \text{HM}\right\}$.
where $\text{X} \in \left\{\text{CT}, \text{CA}, \text{HM}\right\}$. Constraint \eqref{OPP1B} represents the legitimate range of the phase-shifts. Since STARS (not includes controller) is passive devices, their amplitude response is limited by the conservation of energy, as shown in \eqref{OPP1E}. Constraint \eqref{OPP1D} is a QoS constraint specifically the minimum data rate. Constraint \eqref{OPP1D} is a constraint on the maximum caching capacity. Finally, \eqref{OPP1G} is the maximum power constraint for the BS and Caching-at-STARS controller.

According to \textbf{Problem 1}, the optimization problem with coupled T\&R phase-shift can be formulated as \textbf{Problem 2}:
\begin{subequations}
\begin{align}\label{OPP2}
%&\min_{\mathbf{c}_{b,t},\mathbf{c}_{c,t},\mathbf{w}_{f,k_t,t}, \mathbf{w}_{f,k_r,t}, \mathbf{\Theta}_{\mathcal{T},t},\mathbf{\Theta}_{\mathcal{R},t}} \sum_{t=1}^{T} \parallel \mathbf{w}^2_{\mathbf{f},t}\parallel + \lambda_t w_b,\\
%\tilde{\mathcal{P}}: & \min_{\left\{\mathbf{P}^b_{{k,t}}, P_{k,t}^c, \mathbf{\Theta}_{q,t}, \mathbf{c}_{t}^i\right\}} \sum_{t=1}^{T} \left( P_{w,t} + \lambda_{r,t} P_{bh} + \lambda_{u,t} P_{u} \right),\\
\text{P2}: \min_{\mathbf{P}^b, P^c, \mathbf{\Theta}, \mathbf{c}} \quad & \sum_{t=1}^{T} P_{s,t},\\
\textrm{s.t.} \; \ \
%& -\pi \leq \theta_{\mathcal{R},n,t} \le \pi, \forall n, \forall t,\label{OPPC}\\
& cos(\theta_{q,n,t}-\theta_{\overline{q},n,t})=0, \forall q, \forall n, \forall t, \label{OPP2B}\\
& \eqref{OPP1B}-\eqref{OPP1G},
\end{align}
\end{subequations}
where the \eqref{OPP2B} represents the coupled phase-shift constraint. As a consequence, in the following sections, we develop new efficient TD3-based algorithms to solve the above problems.

\section{Frequency-Aware Based TD3 Optimization Design With Independent T\&R Phase-Shift Caching-at-STARS}

As a long-term decision process, our proposed independent T\&R phase-shift Caching-at-STARS optimization problem (P1) obeys a Markov decision process (MDP). For the proposed problem, two key challenges need to be solved. The first challenge is to make appropriate caching placement decisions in the Caching-at-STARS system to reduce signal loss from multiple forwarding. The second challenge is to make appropriate active and passive beamforming coefficients decisions in the Caching-at-STARS system to enhance the quality of the signal for wireless transmission. In this section, we propose a frequency-aware based TD3 (FA-TD3) algorithm to solve these two challenges, simultaneously.

%these two sub-problems are jointly solved as a Markov decision problem with continuity in both the state space and the action space, and a TD3 algorithm is proposed to solve the problem.
\vspace{-0.5cm}
\subsection{Independent T\&R Phase-Shift Caching-at-STARS MDP Model}

In the MDP formalization of DRL, the agent makes action $\mathbf{a}_t \in \mathcal{A}$ decision by sensing the current state $\mathbf{s}_t \in \mathcal{S}$ at time slot $t \in \mathcal{T}$, where $\mathcal{S}$ and $\mathcal{A}$ represent the state space and action space. After the agent has performed the action $\mathbf{a}_t$, it is rewarded with $r_t$ based on immediate feedback, and the environment state is transformed to $\mathbf{s}_{t+1}$ accordingly. A step of MDP is expressed as a Markov transition tuple $<\mathbf{s}_t, \mathbf{a}_t, r_t, \mathbf{s}_{t+1}>$ and recorded in the replay buffer for the agent's training. In the following, we will describe the state space $\mathcal{S}$, action space $\mathcal{A}$, and reward $r_t$ in independent T\&R phase-shift Caching-at-STARS MDP model.

\subsubsection{State Space}
Given that content caching and wireless signal transmission are jointly considered in the independent T\&R phase-shift Caching-at-STARS MDP model. The state for each time slot $t$ includes network caching information, user request information, and global channel state information (CSI), comprising CSI from the BS to Caching-at-STARS, passive transmission from Caching-at-STARS' elements to users, and active transmission from Caching-at-STARS' relay to users. Thus, the state vector at time slot $t$, denoted by $\mathbf{s}_t$, is expressed as follows:
%\begin{align}\label{S}
%\mathbf{s}_t=\{\mathbf{c}_{t-1}^b, \mathbf{c}_{t-1}^c, f_{k_t,t}, f_{k_r,t}, \mathbf{G}_{b,t}, \mathbf{h}_{c,k_r,t}, \mathbf{h}_{c,k_t,t}, h_{c,k_r,t}^d, h_{c,k_t,t}^d \}.
%\end{align}
\begin{align}\label{S}
\mathbf{s}_t=\{\mathbf{c}_{t-1}^i, f_{k,t}, \mathbf{G}_{b,t}, \mathbf{h}_{c,k,t}, h_{c,k,t}^d \}.
\end{align}

\subsubsection{Action Space}

The action designed for the independent T\&R phase-shift Caching-at-STARS MDP model needs to consider all decision variables in Eq. \eqref{OPP1}. The action includes caching replacement vectors for the BS and Caching-at-STARS, a vector that stores the active and passive beamforming coefficients at the BS and Caching-at-STARS, and the power allocation vector for active Caching-at-STARS signal transmission. The decision variables mentioned above are categorized into two groups based on their discrete nature: continuous variables, including $\mathbf{P}^b_{{k,t}}, \mathbf{\Theta}_{\mathcal{J},t}$, and $P_{k,t}^c$, and discrete variables, including $\mathbf{c}_t^i$.

During the cache decision process, directly setting $\mathbf{c}_t^i$ as the cache decision actions is difficult for the the convergence of the algorithm and the maintenance of constrains \eqref{OPP1C}. The reason is that the dimensions of the discrete variables $\mathbf{c}_t^i$ is related to the number of in-network content and the caching capability of nodes, and the caching capacity of the nodes is usually much smaller than the amount of content. The $\mathbf{c}_t^i$ as action will select $C_i$ contents from all the in-network contents for caching, and the huge dimensional difference leads to slow and unstable convergence of algorithm. Moreover, it is difficult for an algorithm to optimize both discrete and continuous variables. To address these challenges, we will serialize the cache decision action and propose two different action policy designs.

\textbf{\textit{Equal-Width Continuous Cache Action Space}}: To reduce the dimension of the cache action and improve the convergence of the algorithm, we will build the cache decision action based on the low-dimensional node caching capability. The new cache decision vector $\mathbf{c}_t^{i,*}$ and cache decision action $\mathbf{a}_t^{\mathbf{c}^{i,*}}$ for node $i$ at time slot $t$ can be expressed as
\begin{align}\label{E-a}
\mathbf{c}_t^{i,*} = [f_{i,t}^1,f_{i,t}^2,...,f_{i,t}^n,...,f_{i,t}^{C_i}] \leftarrow \mathbf{a}_t^{\mathbf{c}^{i,*}} = [\varphi_{i,t}^1,\varphi_{i,t}^2,...,\varphi_{i,t}^n,...,\varphi_{i,t}^{C_i}] ,
\end{align}
where $f_{i,t}^n$ and $\varphi_{i,t}^n$ represent the content number cached at the $n$-th location of the node $i$ at time slot $t$ and the corresponding DRL output. Based on the equal-width continuous policy, $f_{i,t}^n$ can be decoded as
\begin{align}\label{E-a}
f_{i,t}^n= \lceil \frac{\varphi_{i,t}^n \cdot F}{\varphi_{max} - \varphi_{min}} \rceil,
\end{align}
where $\varphi_{max}$ and $\varphi_{min}$ represent the upper and lower bounds for the DRL optimization variables, respectively.

\textbf{\textit{Frequency-Aware Dynamic Continuous Cache Action Space}}: The equal-width policy effectively serialises the cache action space, but also does not take full advantage of the knowledge generated by user requests in the actual physical space. For this reason, a frequency-aware policy is proposed to serialise the cache action space. The frequency information of user requests for content is utilized to dynamically adjust the mapping between $\mathbf{c}_t^{i,*}$ and $\mathbf{a}_t^{\mathbf{c}^{i,*}}$, thus increasing the probability of popular content being cached. In frequency-aware dynamic continuous cache action space, the length of the $n$-th content in the cache decision space is calculated as
\begin{align}\label{E-a}
L_n= \frac{\left(\varphi_{max} - \varphi_{min}\right) \cdot \chi}{F} + \frac{\left(\varphi_{max} - \varphi_{min}\right) \cdot \left(1-\chi\right) \cdot \Psi_n}{\sum_{k=1}^F \Psi_k},
\end{align}
where $\Psi_n$ represents the number of times $k$-th content has been requested, and $\chi$ represents the scale for dynamically adjusting the cache action space. The probability that the system caches $n$-th in-network content is $L_n / (\varphi_{max} - \varphi_{min})$. It can be seen that the utilization of users' content request frequency information results in a higher likelihood of caching popular content.
$f_{i,t}^n$ can be decoded as
\begin{align}\label{E-a}
f_{i,t}^n = \Delta \leftarrow \: \sum_{\delta=1}^{\Delta-1} L_\delta \leq \varphi_{i,t}^n \leq \sum_{\delta=1}^\Delta L_s, \: 1 \leq \Delta \leq F.
\end{align}

Then, the normalized output of the remaining continuous actions generated by actor network may be directly decoded as the continuous decision variables in Eq. \eqref{OPP1}
\begin{align}\label{E-a}
&\{\mathbf{P}^b \} \leftarrow \mathbf{a}_t^{\mathbf{P}^b}, \\
&\{\boldsymbol{\mathbf{\theta}}_{\mathcal{T}}, \boldsymbol{\mathbf{\theta}}_{\mathcal{R}}, \boldsymbol{\mathbf{\beta}}_{\mathcal{T}}, \boldsymbol{\mathbf{\beta}}_{\mathcal{R}} \} \leftarrow \mathbf{a}_t^{\mathbf{\Theta}}, \\
&\{P^c\}  \leftarrow \mathbf{a}_t^{P^c}.
\end{align}

Therefore, all actions in the independent T\&R phase-shift Caching-at-STARS MDP model are continuous, and the entire continuous action can be formulated as
\begin{align}\label{A}
\mathbf{a}_t = \{\mathbf{a}_t^{\mathbf{c}^{i,*}_e} / \mathbf{a}_t^{\mathbf{c}^{i,*}_f}, \mathbf{a}_t^{\mathbf{P}^b}, \mathbf{a}_t^{\mathbf{\Theta}}, \mathbf{a}_t^{P^c}\},
\end{align}
where $\mathbf{a}_t^{\mathbf{c}^{i,*}_e}$ and $\mathbf{a}_t^{\mathbf{c}^{i,*}_f}$ represent equal-width and frequency-aware dynamic continuous cache action, respectively.

\subsubsection{Reward Function}
The ultimate objective of DRL is to discover an optimal path of state transitions that maximizes the accumulated reward obtained by the agent. The design of the reward function directly affects the exploration of problem (P1) and the convergence of the algorithm. In order to jointly optimize content caching and power allocation on the basis of meeting user service requirements \eqref{OPP1D}, these factors will be taken into account in the design of the reward function. Thus, the reward function at time slot $t$ can be formulated as
\begin{align}\label{E-a}
r_t = \sum_{\substack{k \in \mathcal{K} \\ R_{k,t}^{\text{X}} \geq R_\text{QoS}}} r_q - \varpi_p \cdot P_{s,t} + \varpi_h \cdot \sum_{i \in \mathcal{I}} H_t^i.
\end{align}
where $r_q$ represents the reward for meeting user service requirements. $\varpi_p$ and $\varpi_h$ represent the penalty and incentive coefficients for power consumption and cache hit rate in the reward function, respectively. $H_t^i$ is the number of cache hits for node $i$ at time slot $t$.
\vspace{-0.5cm}
\subsection{Training Process of Frequency-Aware Based TD3 Algorithm}

The action space $\mathcal{A}$ of independent T\&R phase-shift Caching-at-STARS MDP model is continuous, whether based on equal-width cache action space or frequency-aware dynamic cache action space. TD3 algorithm, as a branch of the actor-critic architecture DRL, is suitable for addressing continuous control problems. Thus, a variant of TD3, named FA-TD3, will be developed to fully utilize user request frequency information and solve problem (P1) with a continuous decision space.

%As shown in Fig. \ref{fig:FA-TD3},
The FA-TD3 algorithm primarily relies on actor and critic networks for action decision-making and evaluation, respectively. Specifically, the actor networks select actions in a given state by fitting the action decision function $\mu(\mathbf{s}_t|\mathbf{\boldsymbol{\omega}^{\mu}})$ for state $\mathbf{s}_t$ at time slot $t$. On the other hand, the critic networks evaluate the value of action choices in a given state by fitting the state-action value function $\phi(\mathbf{s}_t,\mathbf{a}_t|\mathbf{\boldsymbol{\omega}^{\phi}})$. The parameters of the actor and critic networks are denoted by $\boldsymbol{\omega}^{\mu}$ and $\boldsymbol{\omega}^{\phi}$, respectively.
%The core of the FA-TD3 is the actor networks and critic networks, which are adopted for action decision-making and action evaluation respectively. In other words, the actor networks are responsible for selecting actions in a given state, by fitting the action decision function $\mu(\mathbf{s_t}|\theta^{\mu})$ in state $\mathbf{s_t}$ at time slot $t$. The critic networks are responsible for  evaluating the value of action choices in a given state, by fitting the state-action value functions $Q(\mathbf{s_t},\mathbf{a_t}|\theta^{Q})$. Here, $\theta^{\mu}$ and $\theta^{Q}$ are the network parameters of the actor networks and critic networks.
Moreover, there are three important polices in the FA-TD3: twin critic networks, delay soft update, and target policy smoothing.

%\begin{figure}[!t]
%	\centering	
%    \captionsetup{font={small}}
%	\includegraphics[scale=0.93]{FA-TD3.eps}
%	\caption{Flow diagram of the FA-TD3 algorithm.}
%    \vspace{-1cm}
%	\label{fig:FA-TD3}
%\end{figure}

As mentioned earlier, the ultimate objective of training the FA-TD3 agent is to optimize the decision-making process such that the expected cumulative rewards, denoted by $\mathbb{E}\left[\sum_{i=t}^{T} \gamma^{i-t} \cdot r_i \right]$, are maximized for corresponding action $\mathbf{a_t}$ taken at state $\mathbf{s_t}$, where $T$ represents the final step and $\gamma \in [0,1]$ is the discount factor. According to the Bellman equation, the Q-value of critic network $\phi$ for each action $\mathbf{a_t}$ of the FA-TD3 agent for the state $\mathbf{s_t}$ is
\begin{align}\label{Bellman}
Q_{\phi}(\mathbf{s}_t,\mathbf{a}_t)=\mathbb{E}\left[r(\mathbf{s}_t,\mathbf{a}_t)+ \gamma Q_{\phi}(\mathbf{s}_{t+1},\mathbf{a}_{t+1}) \right].
\end{align}

To overcome the problem of Q value overestimation, double critic networks policy is adopted in FA-TD3 to estimate the actual Q-value, which uses the smaller of the two Q-values to form the targets in the Bellman error loss functions, $y_t$ is calculated by
\begin{align}\label{Y}
y_t = r_t + \gamma \min \left[Q^\prime_{\phi_1}(\mathbf{s}_{t+1},\mu^\prime(\mathbf{s}_{t+1}|\boldsymbol{\omega}^{\mu^\prime}+\epsilon)|\boldsymbol{\omega}^{\phi^\prime_1}),
Q^\prime_{\phi_2}(\mathbf{s}_{t+1},\mu^\prime(\mathbf{s}_{t+1}|\boldsymbol{\omega}^{\mu^\prime}+\epsilon)|\boldsymbol{\omega}^{\phi^\prime_2}) \right],
\end{align}
where $\boldsymbol{\omega}^{\mu^\prime}$ and $\boldsymbol{\omega}^{\phi^\prime}$ represent the parameters of the target actor network and critic network, and $\epsilon \thicksim \mathcal{N}(0,\xi)$ represents Gaussian noise with scale $\xi(\varkappa)$ related to the training episode $\varkappa$. In order to enhance the exploration of the FA-TD3 algorithm, a relatively large value of $\xi(\varkappa)$ is set in the early stage of training, and as the algorithm iterates, $\xi(\varkappa)$ gradually decreases to improve the exploitation of the algorithm. This policy of adding noise to the target action is target policy smoothing, which can make it harder for the policy to overfit to narrow peaks in the value estimate. Thus, the target action can be express as
\begin{align}\label{action_generate}
\mathbf{a^\prime}_{t} = \mu^\prime(\mathbf{s}_{t}|\boldsymbol{\omega}^{\mu^\prime})+\epsilon.
\end{align}

Moreover, it should be noted that the input of the actor network needs to add content request frequency $\Psi_k$ to dynamiclly adjust the action space division scale, and the output of the actor network in FA-TD3 also needs to add noise to avoid the algorithm falling into local optimum.

The experience generated by the algorithm in different training stages is stored in the reply buffer to be sampled to calculate TD-error and then update the critic network. $y_t$ will be used as a label to train the either of the twin critic networks by minimizing the loss function
\begin{align}\label{loss_critic}
L(\boldsymbol{\omega}^{\phi}) = \frac{1}{e} \sum_e \left[y_t - Q_{\phi}(\mathbf{s}_t,\mathbf{a}_t|\boldsymbol{\omega}^{\phi})\right]^2,
\end{align}
where $e$ is the batch size from the reply buffer. The target networks adopt delay soft update policy to iterate their parameters $\mu(\mathbf{s}_t|\omega^{\mu})$. This policy refers that FA-TD3 algorithm updates the target networks less frequently than the critic networks to reduce the oscillation of the algorithm, which can be expressed as
\begin{align}\label{up_target}
\boldsymbol{\omega}^{\mu^\prime} \leftarrow \tau\boldsymbol{\omega}^{\mu} + (1-\tau) \boldsymbol{\omega}^{\mu^\prime}, \\
\boldsymbol{\omega}^{\phi^\prime} \leftarrow \tau\boldsymbol{\omega}^{\phi} + (1-\tau) \boldsymbol{\omega}^{\phi^\prime},
\end{align}
%\begin{align}\label{E-a}
%}
%\end{align}
where $\tau$ is the update coefficient and $0<\tau \ll 1$.

Finally, the actor network is trained by the critic network gradient
\begin{align}\label{up_a}
\nabla_{\boldsymbol{\omega}^\mu}J(\boldsymbol{\omega}^\mu) = \frac{1}{e} \sum_e \nabla_{\mathbf{a}}Q_{\phi_1}(\mathbf{s}_t,\mathbf{a}_t|\boldsymbol{\omega}^{\phi_1})|_{\mathbf{a}_t=\mu(\mathbf{s}_t)} \nabla_{\boldsymbol{\omega}^\mu} \mu(\mathbf{s}_t|\boldsymbol{\omega}^\mu).
\end{align}
%\begin{align}\label{Actor}
%\nabla_{\bm{\theta}^\mu}J = \frac{1}{e}\sum_{e}\nabla_\mathbf{a} Q(\mathbf{s}_e,\mathbf{a}_e|\bm{\omega}^Q)|_{\mathbf{s}_e=\mathbf{s}_t,\mathbf{a}_e=\mu(\mathbf{s}_t)}\nabla_{\bm{\omega}^\mu} \mu(\mathbf{s}_e|\bm{\omega}^\mu|_{\mathbf{s}_e=\mathbf{s}_t}).
%\end{align}

%\begin{align}\label{S}
%\mathbf{s}_t =\{\bm{H}_{b,r,t},\bm{H}_{b,\mathcal{R},t},\bm{H}_{b,\mathcal{T},t},\bm{H}_{r,\mathcal{R},t},\bm{H}_{r,\mathcal{T},t} \}.
%\end{align}
\vspace{-0.5cm}
\subsection{Neural Network Architecture} \label{I_MDP}

To ensure that the neural network in FA-TD3 can accurately fit the selection of action $\mathbf{a}_t$ and the evaluation of Q-value $Q\left(\mathbf{s}_t,\mathbf{a}_t\right)$, the construction of the neural network and the setting of the associated hyperparameters are significant. There are six neural networks in FA-TD3, including the (target) actor network and the (target) twin critic networks. In particular, to facilitate the training of the neural networks, the target networks will have the same structure as its corresponding actor or critic network to serve as the training label for the loss function.

The actor networks include the input layer, fully connected hidden layer, and output layer. The dimension of input and output layer correspond to the dimensions of state $\mathbf{s}_t$ and action $\mathbf{a}_t$, respectively. Moreover, The action space of the independent T\&R phase-shift caching-at-STARS MDP model is continuous, both in the frequency-aware dynamic action space and in the equal-width cache action space. The activation function Tanh is employed in output layer to characterize the continues action. Moreover, three fully connected hidden layers with ReLu as the activation function and containing 64 neurons were used to construct the actor network. The structure of critic networks is similar to that of actor networks. The differences are that the dimension of the input layer of critic networks is equal to the sum of the dimensions of state and action and the output is the Q-value.

\section{Cooperative TD3-DQN Optimization Design With Coupled T\&R Phase-Shift Caching-at-STARS}

In this section, we turn our attention to coupled T\&R phase-shift Caching-at-STARS optimization problem (P2). As a long-term decision process, this problem remains suitable for solving by DRL. However, different from independent T\&R phase-shift Caching-at-STARS optimization problem, coupled optimization problem adds constraint on STARS phase-shift control \eqref{OPP2B}. This constraint is reflected in the coupled T\&R phase-shift Caching-at-STARS MDP model, where for any element $n$ on Caching-at-STARS with a reflection phase-shift $\theta_{\mathcal{R},n}$, the transmission phase-shifts can only be $\theta_{\mathcal{R},n}+\frac{\pi}{2}$ and $\theta_{\mathcal{R},n}-\frac{\pi}{2}$. Therefore, the reflection phase-shift can be treated as a continuous variable optimized within the limits of constraint \eqref{OPP1B}, while the transmission phase-shift is considered as a binary discrete decision variable \eqref{OPP2B}.

Obviously, we can continue to serialize this binary discrete decision variable by equal-width continuation strategy.
%However, the decision to transmission phase-shift $\theta_{\mathcal{T},n}$ is associated with reflection phase-shifts. It would be more effective to use the optimization results of the reflection phase-shift as a priori knowledge to guide the decision to transmit the binary phase-shift.
However, since the decision to transmit phase-shift $\theta_{\mathcal{T},n}$ is related to reflection phase-shifts, it would be more efficient to utilize the optimization results of reflection phase-shifts and caching replacement at current time slot as prior knowledge to guide the decision of transmitting the binary phase-shift. Thus, we explore another promising algorithm that use two DRL agents to optimize continuous and discrete variables jointly. TD3 (contains its variants FA-TD3) and DQN are capable for continuous and discrete action control, respectively, based on their ability to perceive continuous state spaces. A new cooperative TD3-DQN algorithm is proposed to address this problem, where FA-TD3 and DQN co-agents can optimize continuous and low-dimensional discrete variables respectively, but jointly influence the environment. It is important to note that although the roles of the TD3 and DQN agents are described separately, they work collaboratively as a pair of co-agents to solve the problem and influence each other.
\vspace{-0.5cm}
\subsection{Coupled T\&R Phase-Shift Caching-at-STARS MDP Model}

In the coupled T\&R phase-shift Caching-at-STARS MDP model, the entire action space is divided into a continuous action space and a discrete action space, which are solved by FA-TD3 and DQN agents, respectively. However, how these two agents can collaborate to effectively solve the problem remains an open challenge. To simplify the description, we denote continuous and discrete actions as $\mathbf{a}_t^c$ and $\mathbf{a}_t^d$ respectively. Optimizing actions $\mathbf{a}_t^c$ and $\mathbf{a}_t^d$ in parallel achieves the problem solution, but the decision process for $\mathbf{a}_t^d$ is not inherently influenced by $\mathbf{a}_{t}^c$. Therefore, we use FA-TD3 algorithm, suitable for continuous action control, to optimization model and obtain action $\mathbf{a}_t^c$. Then, in the environment changed by action $\mathbf{a}_t^c$, the DQN agent is used to sense the changed environment and make binary decisions for action $\mathbf{a}_t^d$. The joint TD3-DQN algorithmic framework allows the co-agents to influence each other's decisions and jointly optimize the solution.

\begin{figure}[!t]
	\centering	
    \captionsetup{font={small}}
	\includegraphics[scale=0.93]{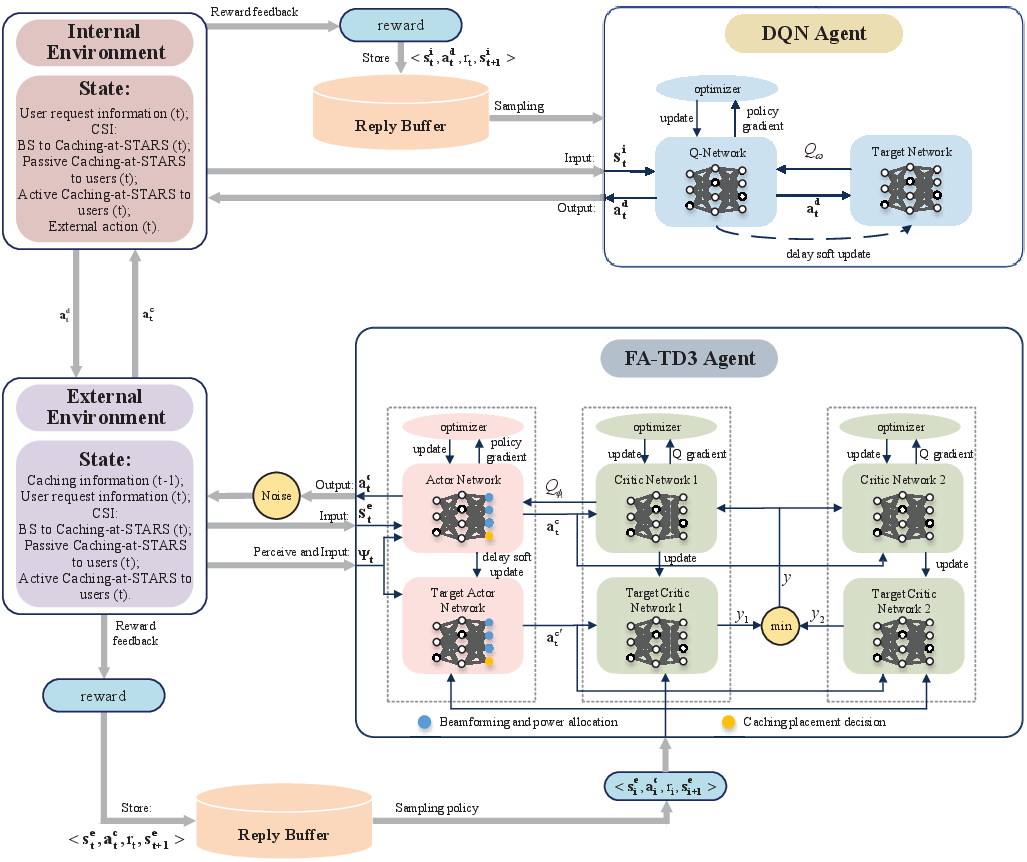}
	\caption{Flow diagram of the cooperative TD3-DQN algorithm.}
    \vspace{-1cm}
	\label{fig:DQN-TD3}
\end{figure}

As shown in Fig. \ref{fig:DQN-TD3}, the environment is divided by the cooperative TD3-DQN algorithm into an external environment and an internal environment. The external environment is the projection of the real network environment, FA-TD3 agent perceives the state of the environment $\mathbf{s}_t^{ex}$ and makes continuous action decisions $\mathbf{a}_t^c$ in time slot $t$. When the external real environment receives the $\mathbf{a}_t^c$ from the FA-TD3 agent, it is transformed into the virtual internal environment. The continuous action $\mathbf{a}_t^c$ will act as a priori knowledge to influence the DQN agent to perceive the environment state $\mathbf{s}_t^{in}$, optimize the action $\mathbf{a}_t^d$ and obtain reward $r_t^{in}$. The action $\mathbf{a}_t^d$ and action $\mathbf{a}_t^c$ in the internal environment state are combined to act on the external environment and produce a reward $r_t^{ex}$.
Therefore, there are two MDPs in cooperative TD3-DQN algorithm based coupled T\&R phase-shift Caching-at-STARS model, $<\mathbf{s}_t^{ex}, \mathbf{a}_t^c, r_t^{ex}, \mathbf{s}_{t+1}^{ex}>$ for external environment and $<\mathbf{s}_t^{in}, \mathbf{a}_t^d, r_t^{in}, \mathbf{s}_{t+1}^{in}>$ for internal environment. The state, action and reward function of two MDPs will be discussed in the following, respectively.

\begin{algorithm}
\caption{Cooperative TD3-DQN algorithm}
\renewcommand{\algorithmicrequire}{\textbf{Input:}}
\renewcommand{\algorithmicensure}{\textbf{Output:}}
\label{TD3DQN}
\begin{algorithmic}[1]
		\REQUIRE The set of user request $\mathbf{f}_{k}$.
		\ENSURE active beamforming vector $\mathbf{P}^b$, active Caching-at-STARS transmission power $P^c$, STARS' coefficient $\mathbf{\Theta}$, and content caching replacement vector $\mathbf{c}$.
        \STATE Initialize the network environment, the number of algorithm training period episode $max_{e}$ and step $max_{t}$, and replay buffers in external and internal environments.
        \STATE Initialize the FA-TD3 agent with the actor and target actor networks with parameters $\boldsymbol{\omega}^{\mu}$ and $\boldsymbol{\omega}^{{\mu}^\prime}$, twin critic and target critic networks with parameters $\boldsymbol{\omega}^{\phi_1}$, $\boldsymbol{\omega}^{{\phi}_1^\prime}$, $\boldsymbol{\omega}^{\phi_2}$, and $\boldsymbol{\omega}^{{\phi}_2^\prime}$, where $\boldsymbol{\omega}^{\mu} = \boldsymbol{\omega}^{{\mu}^\prime}$, $\boldsymbol{\omega}^{\phi_1} = \boldsymbol{\omega}^{{\phi}_1^\prime}$, and $\boldsymbol{\omega}^{\phi_2} = \boldsymbol{\omega}^{{\phi}_2^\prime}$;
        \STATE Initialize the DQN agent with the Q-network $\mathbf{\omega}^{\zeta}$ and target Q network $\mathbf{\omega}^{\zeta^\prime}$, where $\boldsymbol{\omega}^{\zeta} = \boldsymbol{\omega}^{\zeta^\prime}$;
        \FOR{episode $= 1 \rightarrow max_{e}$}
        \STATE Reinitialize the network environment to $\mathbf{s}_{t=0}$
        \FOR{step $t = 1 \rightarrow max_{t}$}
        \STATE FA-TD3 agent perceives external environment $\mathbf{s}_{t}^{ex}$ and chooses action $\mathbf{a}^c_{t}$ with \eqref{action_selection_ex};
        \STATE DQN agent perceives external environment $\mathbf{s}_{t}^{in}$ and chooses inner action $\mathbf{a}^d_{t}$ with \eqref{action_selection_in_train};
        \STATE Execute action $\mathbf{a}^d_{i,t}$ in the inner environment;
        \STATE Calculated the reward $r_{t}^{in}$ and perceive the next state $\mathbf{s}_{t+1}^{in}$;
        \STATE Record $e \{\mathbf{s}_{t}^{in},\mathbf{a}^c_{i,t},r_{t}^{in},\mathbf{s}_{t+1}^{in}\}$;
        \STATE Sample a batch of record $e$ from DQN replay buffer;
        \STATE Train Q-network parameters $\mathbf{\omega}^{\zeta}$ with \eqref{loss_DQN};
        \STATE DQN agent choose outer action $\mathbf{a}^d_{e,t}$ with \eqref{action_selection_in_output};
        \STATE Execute ${\mathbf{a}^d_{e,t}, \mathbf{a}^c_{t}}$ in the external environment;
        \STATE Calculated the reward $r_{e,t}$ and perceive the next state $\mathbf{s}_{t+1}^{ex}$;
        \STATE Record $e \{\mathbf{s}_{t}^{ex},\mathbf{a}_{t}^c,r_{t}^{ex},\mathbf{s}_{t+1}^{ex}\}$ in TD3 replay buffer;
        \STATE Sample a batch of record $e$ from TD3 replay buffer;
        \STATE Calculate target (labal) according to \eqref{Y};
        \STATE Train twin critic network with a gradient descent step \eqref{loss_critic};
        \STATE Train actor network with \eqref{up_a};
        \STATE Update the target networks with \eqref{up_target};
        \STATE $\mathbf{s}_{t}^{ex} \leftarrow \mathbf{s}_{t+1}^{ex}$;
        \ENDFOR
        \ENDFOR

\end{algorithmic}
\end{algorithm}

\vspace{-0.5cm}
\subsection{External MDP Model and FA-TD3 Agent} \label{CE_MDP}

\subsubsection{State Space}
The external MDP model still needs to be optimized for caching and beamforming, so the FA-TD3 agent needs to perceive network caching information, user request information, and global CSI. The state of external MDP model at slot $t$ can be expressed as
\begin{align}\label{S_ex}
\mathbf{s}_t^{ex}=\{\mathbf{c}_{t-1}^i, f_{k,t}, \mathbf{G}_{b,t}, \mathbf{h}_{c,k,t}, h_{c,k,t}^d \}.
\end{align}

\subsubsection{Action Space}
The external MDP model only optimizes continuous variables and its action does not contain control variables of transmission phase shift. The action of the external MDP model at slot $t$ can be expressed as
\begin{align}\label{A_ex}
\mathbf{a}_t^c = \{\mathbf{a}_t^{\mathbf{c}^{i,*}_f}, \mathbf{a}_t^{\mathbf{P}^b}, \mathbf{a}_t^{\mathbf{\boldsymbol{\theta}_\mathcal{R}}}, \mathbf{a}_t^{\mathbf{\boldsymbol{\beta}}}, \mathbf{a}_t^{P^c}\},
\end{align}
where the caching decision variable is denoted by $\mathbf{a}_t^{\mathbf{c}^{i,*}_f}$ since the FA-TD3 is employed in the optimization of external MDP model.

Three action selection policies are employed in the cooperative DQN-TD3 algorithm. In the early stages of FA-TD3 training, noise policy is adopted to improve the exploration, and optimal policy is used to improve the exploitation in the later training stages. This policy can be formulated as
\begin{align}\label{action_selection_ex}
\mathbf{a}_t^c=
\begin{cases}
\mu(\mathbf{s}_{t}^{ex}|\boldsymbol{\omega}^{\mu})+\epsilon, &  t \leq t_o, \\
\mathop{\arg\max}\limits_{\mathbf{a}_{t}^c} \left[Q(\mathbf{s}_{t}^{ex},\mathbf{a}_{t}^c | \boldsymbol{\omega}^{\phi_1}), Q(\mathbf{s}_{t}^{ex},\mathbf{a}_{t}^c | \boldsymbol{\omega}^{\phi_2})\right],&   t > t_o,
\end{cases}
\end{align}
where $t_o$ is the division threshold between the early and later training stages of the algorithm. Moreover, the remaining action selection policy will be introduce in action selection of internal environment.

\subsubsection{Reward Function}
The reward function of external MDP model is set up on the similar principles as coupled T\&R phase-shift Caching-at-STARS MDP model, it can be expressed as
\begin{align}\label{r_ex}
r_t^{ex} = \sum_{\substack{k \in \mathcal{K} \\ R_{k,t}^{\text{X}} \geq R_\text{QoS}}} r_q - \varpi_p \cdot P_{s,t} + \varpi_h \cdot \sum_{i \in \mathcal{I}} H_t^i.
\end{align}

\begin{remark}\label{Remark1}
In the cooperative TD3-DQN algorithm, the internal virtual environment is nested within the external environment. The internal MDP model takes the continuous action $\mathbf{a}_t^c$ obtained from the external environment optimization as part of the state, which can be used to optimize discrete actions $\mathbf{a}_t^d$. The continuous and discrete actions are then integrated to act jointly on the external environment and obtain external reward $r_t^{ex}$. Therefore, the reward function of the external environment needs to be set taking into account both the user demand satisfaction status, the network node cache status and the network power consumption status.
\end{remark}

\subsubsection{Neural Network Architecture and Training Process}

The neural network architecture and its training process of FA-TD3 agent in external MDP model of coupled T\&R phase-shift Caching-at-STARS model are similar to those of the independent T\&R phase-shift Caching-at-STARS model. However, it is important to highlight that the output layer dimension of the actor network in the FA-TD3 agent has been modified to match its corresponding continuous action dimension.
\vspace{-1cm}
\subsection{Internal MDP Model and FA-TD3 Agent} \label{CI_MDP}

\subsubsection{State Space}
In internal MDP model, the action $\mathbf{a}_t^c$ obtained from the external environment will be treated as the priori knowledge to assist the training of the internal MDP model. Therefore, the internal MDP model can be express as
\begin{align}\label{S_in}
\mathbf{s}_t^{in}=\{f_{k,t}, \mathbf{G}_{b,t}, \mathbf{h}_{c,k,t}, h_{c,k,t}^d, \mathbf{a}_t^{ex} \},
\end{align}
where the caching state of internal MDP model is $\mathbf{c}_t^{i}$ in $\mathbf{a}_t^c$ instead of $\mathbf{c}_{t-1}^{i}$ in $\mathbf{s}_t^{ex}$. The updated cache information can assist the DQN agent to make more accurate transmission phase-shift decisions.

\subsubsection{Action Space}
The purpose of DQN agent training is to find a suitable transmission phase-shift control scheme in the current state, it can be expressed as
\begin{align}\label{A_in}
\mathbf{a}_t^d = \{\mathbf{a}_t^{\mathbf{\boldsymbol{\theta}_\mathcal{T}}}\}.
\end{align}

Meanwhile, since DQN is good at making decisions about discrete actions, in other words, it selects the appropriate one from multiple candidate actions. There are many elements in Caching-at-STARS, so the binary code is used to make decisions about the transmission phase-shift. The variable $\mathbf{\boldsymbol{\theta}_\mathcal{T}}$ is converted to a binary string of length $N$, where $0 \leq \mathbf{\boldsymbol{\theta}_\mathcal{T}} \leq 2^N$, and $\mathbf{\boldsymbol{\theta}_\mathcal{T}} \in \mathbb{N}$.

The $\epsilon$-greedy policy is used in the training stage of internal environment to improve the exploration ability of the DQN agent, it can be expressed as
\begin{align}\label{action_selection_in_train}
\mathbf{a}_{i,t}^d=
\begin{cases}
\text{random action}, &  \epsilon, \\
\mathop{\arg\max}\limits_{\mathbf{a}_{i,t}^d} Q(\mathbf{s}_{t}^{in},\mathbf{a}_{i,t}^d | \boldsymbol{\omega}^{\zeta}) ,&  1-\epsilon.
\end{cases}
\end{align}
Then, the optimal policy is used to select the optimal action $\mathbf{a}_{e,t}^d$, which is combined with action $\mathbf{a}_{t}^c$ to enhance the external environment reward. It can be express as
\begin{align}\label{action_selection_in_output}
\mathbf{a}_{e,t}^d=
\mathop{\arg\max}\limits_{\mathbf{a}_{e,t}^d} Q(\mathbf{s}_{t}^{in},\mathbf{a}_{e,t}^d | \boldsymbol{\omega}^{\zeta}).
\end{align}

\subsubsection{Reward Function}
The training process of internal MDP model only need to consider user satisfaction and the system power consumption, so cache hit rate will be ignore in the reward function of internal MDP model. The reward function of internal MDP model can be expressed as
\begin{align}\label{r_in}
r_t^{in} = \sum_{\substack{k \in \mathcal{K} \\ R_{k,t}^{\text{X}} \geq R_\text{QoS}}} r_q - \varpi_p \cdot P_{s,t}.
\end{align}

\subsubsection{Neural Network Architecture and Training Process}
DQN, as a value-based DRL, makes action decisions by estimating action state Q values from the Bellman equation. As shown in Fig. \ref{fig:DQN-TD3}, there are two neural networks in DQN agent, called Q-network and the target network, respectively. The Q-network is used to generate action $\mathbf{a}_t^d$ and the target network is employed to evaluate Q value. In the internal environment, the Q-value can be formulated by
\begin{align}\label{Bellman_DQN}
Q(\mathbf{s}_t^{in},\mathbf{a}_t^d)=\mathbb{E}\left[r(\mathbf{s}_t^{in},\mathbf{a}_t^d)+ \gamma Q(\mathbf{s}_{t+1}^{in},\mathbf{a}_{t+1}^d) \right].
\end{align}

Different form TD3 algorithm, DQN only have a group of Q-network and target network. The maximized Q value generated by the target network will be used as the label for the Q-network training. $e$ MDP quaternions will be sampled from the reply buffer to be used as training data for the Q-network update, the corresponding loss function is denoted as
\begin{align}\label{loss_DQN}
L(\boldsymbol{\omega}^{\zeta}) = \frac{1}{e} \sum_e \left[r_{t}^{in} + \gamma \max_{\mathbf{a}_{t+1}^d} Q^\prime(\mathbf{s}_{t+1}^{in},\mathbf{a}_{t+1}^d | \boldsymbol{\omega}^{\zeta^\prime}) -  Q(\mathbf{s}_t,\mathbf{a}_t|\boldsymbol{\omega}^{\zeta})\right]^2.
\end{align}
where Q-network and target network have the same network structure and different network parameters. Every fixed steps, the parameters of the target network $\boldsymbol{\omega}^{\zeta^\prime}$ are updated by copying the parameters of the Q-network $\boldsymbol{\omega}^{\zeta}$.

\section{Numerical Results}
%
%\subsection{Simulation Settings}

In this section, the simulation results obtained from FA-TD3 algorithm and cooperative TD3-DQN algorithm are provided to evaluate the performance of the proposed Caching-at-STARS system. We assume that the reference locations of BS and RIS (STARS) are set at (150, 0, 15) meters and (0, 150, 5) meters. The $\mathcal{T}$ and $\mathcal{R}$ users are randomly distributed in a circle region at RIS (STARS) with a radius of 3m, and are located on both sides of the RIS (STARS), respectively. All channels are follow the Rician channel model, any channel $\mathbf{\Lambda}$ can be modelled as $\mathbf{\Lambda} = \sqrt{\frac{\rho_0}{d^\vartheta}}\left(\sqrt{\frac{\varepsilon}{1+\varepsilon}} \mathbf{\Lambda}^{\text{Los}} +  \sqrt{\frac{\varepsilon}{1+\varepsilon}} \mathbf{\Lambda}^{\text{NLos}} \right)$,
where $d$ represents the distance between channel nodes, $\vartheta$ represents the path loss exponents, $\rho_0$ represents the path loss at the reference distance of 1m, $\varepsilon$ represents the Rician factor, and $\mathbf{\Lambda}^{\text{Los}}$ and $\mathbf{\Lambda}^{\text{NLos}}$ represent the Los path and random non-Los path.
The time interval between each user request is 20s, while the channel block fading envelope changes once per second. The system parameters is shown in Table \ref{SP}. The performance of \textbf{{Caching-at-STARS}} is compared to the following model:
\begin{itemize}
\item \textbf{{Caching-at-RIS}}: In the whole system, the remote server, BS, and double-spliced RIS have caching capacity, and double-spliced RIS is used to assist wireless communication. The double-spliced RIS is formed by splicing together a pair of RISs that face in opposite directions, and each RIS contains $N/2$ elements.
\item \textbf{{STARS-Aided Edge Caching}}: The remote server and BS have caching capacity, and STARS is used to assist wireless communication.
\item \textbf{{RIS-Aided Edge Caching}}: The remote server and BS have caching capacity, and a double-spliced RIS is used to assist wireless communication \cite{Chen2021TWC}.
\item \textbf{{STARS without Edge Caching}}: In the whole system, only the remote server stores all in-network contents, and STARS is used to assist wireless communication.
\item \textbf{{RIS without Edge Caching}}: The entire system architecture only utilizes a remote server to store all in-network contents, and wireless communication is assisted by a double-spliced RIS.
\end{itemize}

\begin{table*}[t!]
 \tiny
 \caption{System Parameters}\label{SP}
 \centering
 \footnotesize
 \renewcommand\arraystretch{1.2}
 \begin{tabular}{|c|c|c|c|c|c|}
  \hline
  Parameter & Description & Value &   Parameter & Description & Value \\
  \hline
  $M$ & BS antenna number & 4 & $N$ & STARS element number   & 16  \\
  \hline
  $P_\text{max}$ & maximum power per antenna & 20 dBm & $B$ & bandwidth  & 1 MHz  \\
  \hline
  $\varepsilon$& Rician factors & 3dB & $\sigma$& noise power density & -95.2 dBm/MHz \\
  \hline
  $\vartheta$& path loss exponents & 2 & $F$& In-network content number & 1000 \\
  \hline
  $C_b, C_c$ & Maximum caching capacity & 5, 10 & $\alpha$ & Zipf skewness factor & 0.8\\
  \hline
 \end{tabular}
 \vspace{-1cm}
\end{table*}

For the proposed FA-TD3 and cooperative TD3-DQN algorithm, the ``Adam'' optimizer is employed to train the neural network. The default learning rate of all agents are $3\text{e}^{-4}$. All neural network architectures are introduced in Section \ref{I_MDP}, Section \ref{CE_MDP}, and Section \ref{CI_MDP}. The maximum training period episode $e_{max}$ is 1000, and the maximum training period step $t_{max}$ is 100. The reply buffer in FA-TD3 and DQN can collect $1\text{e}^{5}$ and $1\text{e}^{4}$ records. The sampling batch size is 64, which is used for the training of the neural networks. In FA-TD3 algorithm, the cycle of delay soft update is 3. However, in DQN, the Q-network copies the parameters to the target network every two episodes. In noise action selection policies, $\xi(0) = 0.4$ and $\xi(1000) = 0.2$, respectively. The $\epsilon$ is set 0.15 in $\epsilon$-greedy action selection policy to prevent the DQN agent from falling into local optimum. In reward function of the Caching-at-STARS MDP model, the reward for meeting user service requirements $r_q = 1.7$, power consumption penalty coefficient $\varpi_p = 1$ and cache hit rate incentive coefficient $\varpi_p = 3$.

\begin{figure*}[htbp]
\centering
\subfigure[]{
    \label{fig:model_indepented}
    \includegraphics[scale=.47]{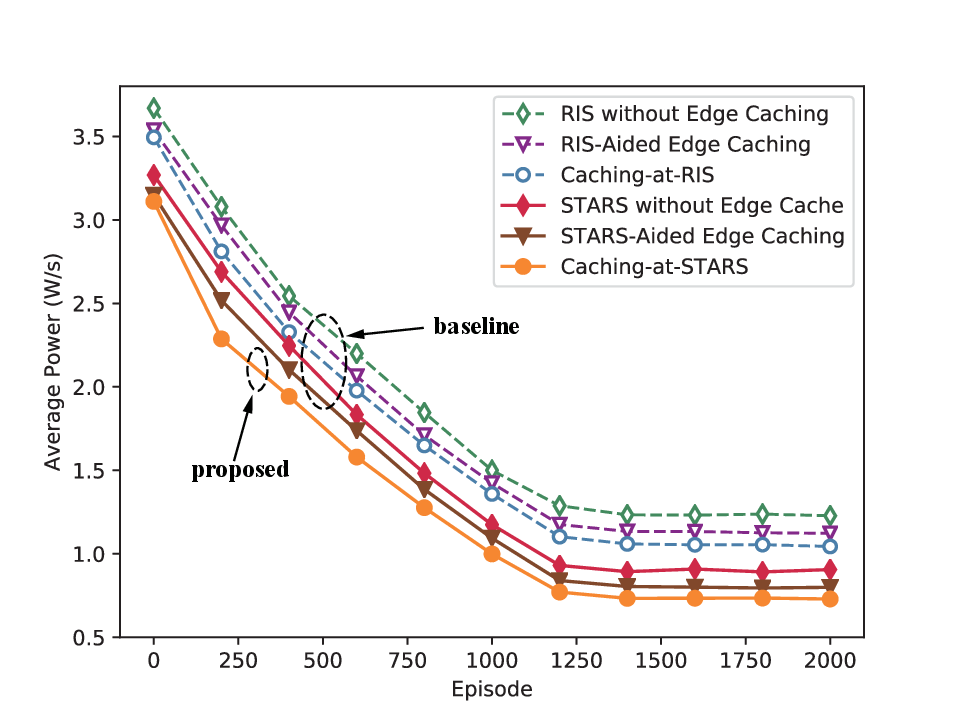}}
%\hspace{0.1in}
%\hfil
\subfigure[]
    {\label{fig:model_coupled}
    \includegraphics[scale=.47]{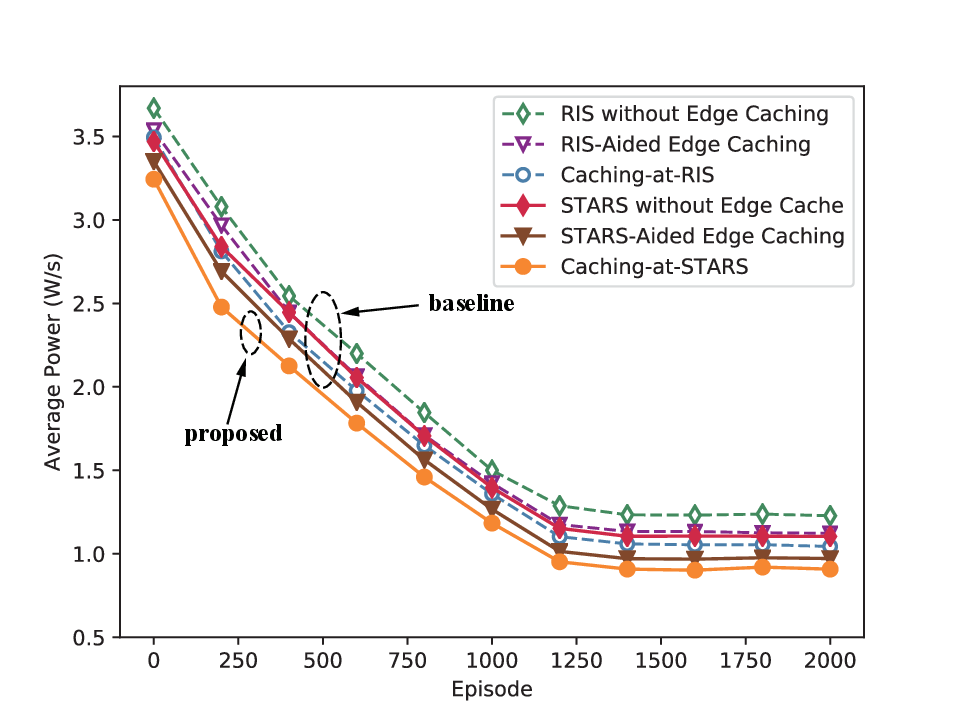}}
\captionsetup{font={small}}
\caption{Power consumption between different STARS (RIS) edge caching models: (a) independent Caching-at-STARS model; (b) coupled Caching-at-STARS model.}
\vspace{-1cm}
\label{fig:power_vs_model}
\end{figure*}

In Fig.~\ref{fig:power_vs_model}, we examine the power consumption between different edge caching models. It can be seen that whether it is an independent ``Caching-at-STARS'' or a coupled ``Caching-at-STARS'', its power consumption is significantly lower than that of the RIS assisted edge caching (including ``Caching-at-RIS'' and ``RIS-Aided Edge Caching''). Among them, the independent and coupled ``Caching-at-STARS'' save power consumption by 12.94\% and 30.11\% compared with ``Caching-at-RIS''. This phenomenon verifies the conclusion of STARS, that is, compared with double-spliced RIS, STARS has higher multipath gain, which leads to stronger signal enhancement for users. Furthermore, ``Caching-at-STARS'' outperforms ``STARS-Aided Edge Caching'' and ``STARS without edge caching'', due to ``Caching-at-STARS'' has a greater probability of satisfying user requests on STARS that is closer to the user, thus reducing path loss and hops. Comparing Fig.~\ref{fig:model_indepented} and Fig.~\ref{fig:model_coupled}, it can be seen that the independent ``Caching-at-STARS'' consumes less power than the coupled ``Caching-at-STARS'', which proves the negative impact of coupling phase-shift control on STARS performance in practical scenarios.

%\begin{figure*}[htbp]
%\centering
%\subfigure[]{
%    \label{fig:model_indepented}
%    \includegraphics[scale=.47]{independent_vs_model_revise.eps}}
%%\hspace{0.1in}
%%\hfil
%\subfigure[]
%    {\label{fig:model_coupled}
%    \includegraphics[scale=.47]{coupled_vs_model_revise.eps}}
%\captionsetup{font={small}}
%\caption{Power consumption between different STAR-RIS caching model: (a) independent Caching-at-STARS; (b) coupled Caching-at-STARS model.}
%\label{fig:0.5cachesize}
%\end{figure*}

\begin{figure*}[htbp]
\centering
\subfigure[]{
    \label{fig:algorithm_indepented}
    \includegraphics[scale=.47]{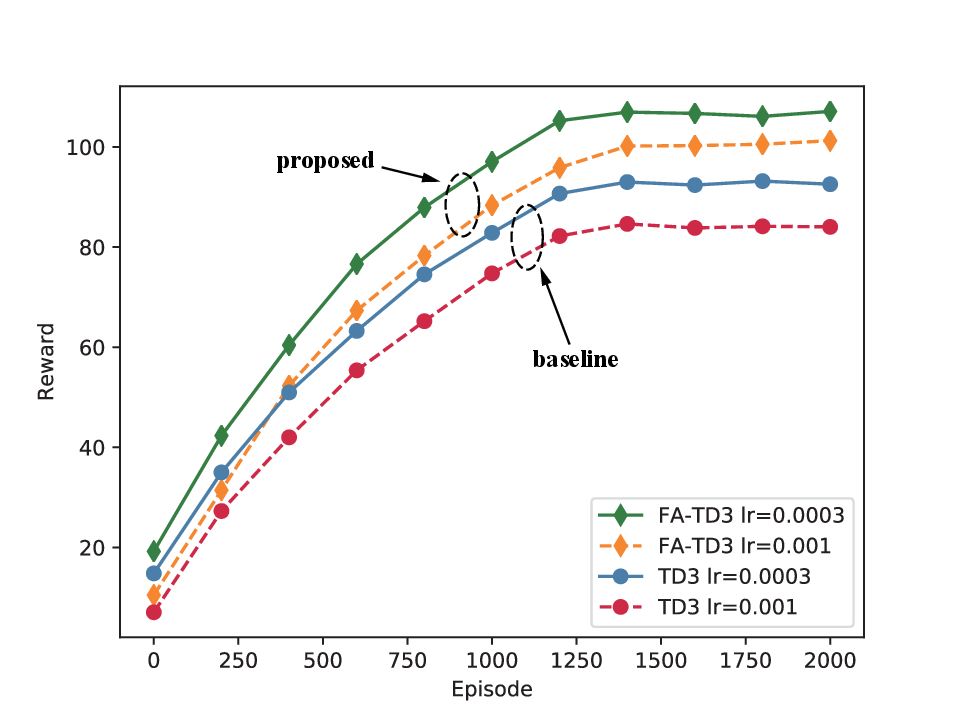}}
%\hspace{0.1in}
%\hfil
\subfigure[]
    {\label{fig:algorithm_coupled}
    \includegraphics[scale=.47]{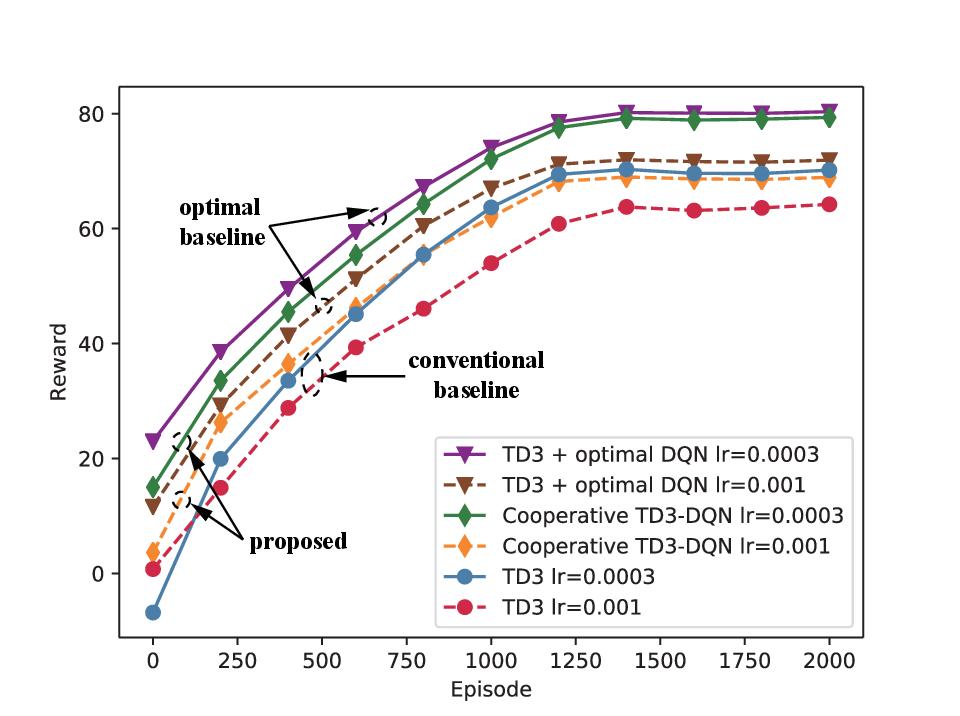}}
\captionsetup{font={small}}
\caption{Reward of different algorithms with various learning rate: (a) independent Caching-at-STARS; (b) coupled Caching-at-STARS model.}
\vspace{-0.8cm}
\label{fig:reward_vs_algorithm}
\end{figure*}

Fig.~\ref{fig:reward_vs_algorithm} presents the reward obtained by the different algorithms with various learning rate. It can be proved that whether it is the traditional TD3, the proposed FA-TD3 or the cooperative TD3-DQN, the algorithm can converge when choosing an appropriate learning rate when optimizing the independent and coupled ``Caching-at-STARS''. In Fig.~\ref{fig:algorithm_indepented}, the gap between FA-TD3 and traditional TD3 with same learning rate gradually increase. The reason is that the frequency-aware dynamic continuous strategy needs to gradually perceive the popularity of the content according to the user's historical requests, so the scale division of the cache decision in the frequency-aware continuous caching action space will gradually become more accurate. In Fig.~\ref{fig:algorithm_coupled}, the rewards of all algorithms are smaller than those in Fig.~\ref{fig:algorithm_indepented} since the self-limitation of coupled STARS phase-shift. In addition, the performance of cooperative TD3-DQN is better than that of traditional TD3, which shows that the scheme of using FA-TD3 and DQN co-agents and controlling the transmission phase-shift after sensing the reflection phase-shift information is effective, though the complexity of the algorithm is increased.

\begin{figure}[h]
	\centering	
    \captionsetup{font={small}} \includegraphics[scale=0.6]{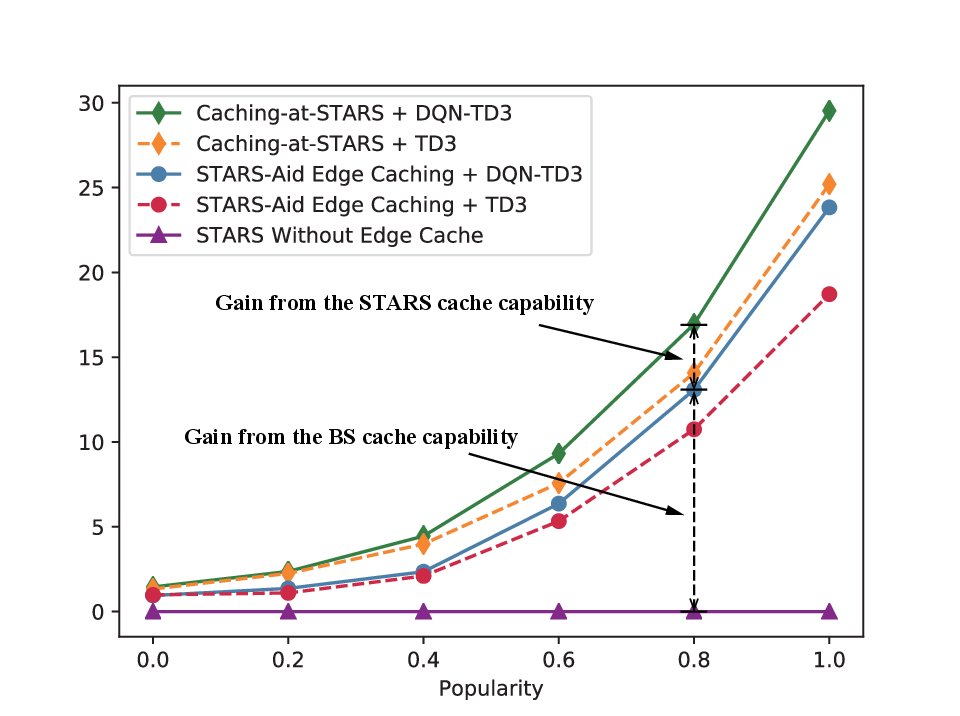}
	\caption{Cache hit rate versus Zipf skewness factor.}
	\label{fig:chr_vs_zipf}
\vspace{-0.8cm}
\end{figure}

Fig.~\ref{fig:chr_vs_zipf} illustrates the cache hit rate of all schemes for various Zipf skewness factor. Here ``Caching-at-STARS'' refers to coupled ``Caching-at-STARS''. Overall, all schemes except ``Without edge cache'' can hit more user requests as the Zipf coefficient increases. A larger Zipf skewness factor means that the content requested by users in the network is more concentrated, so all edge network nodes are more likely to hit user requests in an scenario with a larger Zipf skewness factor. Moreover, the gap between cache hit rate of the same edge caching model solved by the cooperative TD3-DQN algorithm and the traditional TD3 algorithm gradually widens with the increase of the Zipf skewness factor. The reason is that the effect of frequency-aware dynamic continuous strategy is more obvious in an environment where the number of times content is requested differs greatly.
The gain from STARS caching capability is less than the gain of BS capability since the cache capacity of BS is always twice the capacity of the STARS.
As there is no caching capability at the network edge in the "Without Edge Cache" scheme, all requests must be fetched from the remote server, resulting in a cache hit rate of 0.

\begin{figure}[h]
	\centering	
    \captionsetup{font={small}}
	\includegraphics[scale=0.6]{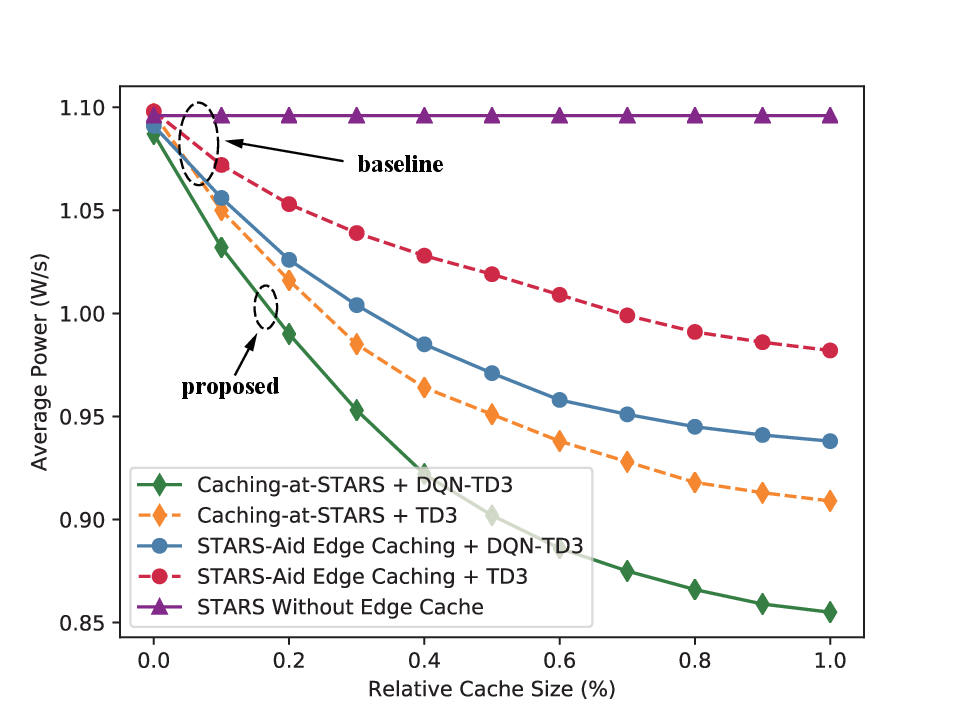}
	\caption{Power consumption versus relative cache size.}
	\label{fig:power_vs_cachesize}
\vspace{-0.8cm}
\end{figure}

Fig.~\ref{fig:power_vs_cachesize} illustrates the power consumption of all schemes for various relative cache size. In this scenario, relative cache size refers to the ratio of STARS cache capacity to the entire network content catalog. Since STARS in ``STARS-Aided Edge Caching'' has no caching capability, the relative cache size of the BS is always twice that of STARS. As the relative cache size of network edge nodes increases, more requests can be satisfied on STARS and BS, so the power consumption of all schemes except ``STARS Without Edge Cache'' gradually decreases. From the perspective of the model, the power consumption of ``Caching-at-STARS'' with cooperative DQN-TD3 is reduced by 9.71\% and 28.19\% compared with that of ``STARS-Aided Edge Caching'' and ``STARS Without edge cache''. From the perspective of the algorithm, cooperative DQN-TD3 solving ``Caching-at-STARS'' and ``STARS-Aided Edge Caching'' has improved by 7.31\% and 4.69\% compared to TD3. The performance of "STARS Without Edge Cache" remains constant since no caching capability is available at its edge nodes.

\section{Conclusion}

A Caching-at-STARS-enabled edge system was investigated, where a network power consumption minimization problem has been formulated to jointly optimize the caching replacement and information-centric wireless signal control. In addition, for independent and coupled T\&R phase-shift models for the Caching-at-STARS, the FA-TD3 and cooperative TD3-DQN algorithms were proposed, which adopted the ideas of serializing discrete variables and directly optimizing continuous and discrete variables, respectively. The numerical results confirmed that the edge caching system enabled by the Catching-at-STARS exhibited more significant advantages in scenarios with large Zipf skewness factor and cache capacity. Furthermore, compared with RIS, STARS is capable of providing services on both sides, and this multipath gain can improve the reuse rate of edge caching. Finally, the proposed FA-TD3 and cooperative TD3-DQN algorithms outperform the conventional TD3 algorithm on Caching-at-STARS problems with different discrete and continuous decision spaces, respectively.

\small
\bibliography{Caching-AT-STARS}
\bibliographystyle{IEEEtran}

\end{document}